\documentclass[aps,pra,10pt, twocolumn,superscriptaddress,amsmath,amssymb]{revtex4-1}
\usepackage[english]{babel}
\usepackage{graphicx}  
\usepackage{epstopdf}
\usepackage{dcolumn}   
\usepackage{enumerate} 
\usepackage{physics} 
\usepackage{bm}        
\usepackage{verbatim}   
\usepackage{color}
\usepackage[dvipsnames]{xcolor}

\begin{document}


\title{Non-asymptotic analysis of quantum metrology protocols beyond the Cram\'{e}r-Rao bound}
\author{Jes\'{u}s Rubio}
\email{J.Rubio-Jimenez@sussex.ac.uk}
\affiliation{Department of Physics and Astronomy, University of Sussex, Brighton BN1 9QH, UK}
\author{Paul Knott}
\affiliation{Centre for the Mathematics and Theoretical Physics of Quantum Non-Equilibrium Systems (CQNE), School of Mathematical Sciences, University of Nottingham, University Park, Nottingham NG7 2RD, UK}
\author{Jacob Dunningham}
\affiliation{Department of Physics and Astronomy, University of Sussex, Brighton BN1 9QH, UK}
   

\begin{abstract}

Many results in the quantum metrology literature use the Cram\'{e}r-Rao bound and the Fisher information to compare different quantum estimation strategies. However, there are several assumptions that go into the construction of these tools, and these limitations are sometimes not taken into account. While a strategy that utilises this method can considerably simplify the problem and is valid asymptotically, to have a rigorous and fair comparison we need to adopt a more general approach. In this work we use a methodology based on Bayesian inference to understand what happens when the Cram\'{e}r-Rao bound is not valid. In particular we quantify the impact of these restrictions on the overall performance of a wide range of schemes including those commonly employed for the estimation of optical phases. We calculate the number of observations and the minimum prior knowledge that are needed such that the Cram\'{e}r-Rao bound \textcolor{black}{is a valid approximation}. Since these requirements are state-dependent, the usual conclusions that can be drawn from the standard methods do not always hold when the analysis is more carefully performed. These results have important implications for the analysis of theory and experiments in quantum metrology.
 
\end{abstract}

\maketitle


\section{Introduction}

Quantum metrology employs quantum resources to enhance the estimation of unknown parameters of interest that are not directly measurable \cite{giovanetti2006review, Dunningham2006, Giovannetti2011, Rafal2015}. Its final aim is to find the strategy that can extract information with the greatest possible precision for a given amount of physical resources, and thus it is an optimization problem. To solve it, first we need to define a mathematical quantity that acts as a figure of merit and informs us about the error of the estimation process. We then minimize that quantity with respect to the elements that we can typically control, that is, the physical state of the system, the measurement scheme and the statistical functions employed in the analysis of the experimental data. 

A widely used method to compare estimation schemes consists in minimizing the mean square error by approaching the Cram\'{e}r-Rao bound, where the latter is defined in terms of the Fisher information \cite{Helstrom1976, kay1993, cox2000}. Although this procedure has its merits and significantly simplifies the analysis of a given strategy, in general it is only suitable when the available prior knowledge is enough to adopt a local approach and the number of experimental observations is asymptotically large \cite{Rafal2015, paris2009, jarzyna2015true,Jaynes2003}. The latter limitation has been addressed in the context of the maximum-likelihood strategy \cite{braunstein_maxlikelihood1992, braunstein_gaussian1992}, and more recently with the quantum Ziv-Zakai and Weiss-Weinstein bounds \cite{tsang2012, tsang2016}, which also incorporate the effect of the prior information. Nevertheless, the previous restrictions are somestimes not taken into account, in spite of the fact that a naive use of the Fisher information can predict schemes with an apparent infinite precision \cite{AlfredoRivas2012, berry2012infinite, zhang2013} which are inefficient in practice \cite{tsang2012, giovannetti2012subheisenberg, berry2012infinite, Pezze2013, Rafal2015}. Since in general it is not possible to foresee when and how the Cram\'{e}r-Rao bound is going to fail in a concrete practical scenario from the asymptotic theory itself, a closer analysis of those schemes that are asymptotically optimal is needed.

The aim of this work is to investigate the regime of validity of the quantum Cram\'{e}r-Rao bound for specific strategies that are commonly employed in the context of quantum metrology. Moreover, we provide quantitative results to understand what happens in practice with the conclusions extracted from the Cram\'{e}r-Rao bound in the regime where it is not a valid approximation. This is achieved by utilising a versatile numerical framework that combines different known Bayesian techniques in a pragmatic way to answer the following question: if we have designed a quantum experiment using the criteria of the asymptotic theory, what is the impact of this simplification on the overall performance when the number of observations is not large enough?

The paper is organised as follows. We start by reviewing the Cram\'{e}r-Rao bound as an asymptotic approximation for the Bayesian error and the basic tools of quantum estimation theory in Section \ref{theory}. Section \ref{method} develops the methodology that we have followed, and our main results are presented and discussed in Section \ref{results}. In particular, we have selected several states commonly used in optical interferometry and we have obtained the mean square error for an asymptotically optimal scheme. This gives us the exact value of the uncertainty for any number of observations. Secondly, we have studied the deviations from the asymptotic approximation and the number of observations needed such that the relative error between the quantum Cram\'{e}r-Rao bound and the exact Bayesian error is small. In addition, we have shown that the numerical approximation of the exact calculation is consistent with the quantum Ziv-Zakai and Weiss-Weinstein bounds. 

Our results verify that both the number of observations and the minimum prior knowledge needed to achieve the asymptotic regime are state-dependent. This has allowed us to show how the conclusions about the relative performance of different states change in the non-asymptotic regime for optical schemes. As a consequence, in general we can say that maximizing the Fisher information alone does not always guarantee the best precision for experiments with a limited number of observations.

\section{Basic theory} \label{theory}

This section includes a summary of the context needed to understand in which sense the Cram\'{e}r-Rao bound can be seen as an approximation and how this motivates our analysis. A more comprehensive introduction to estimation theory and its application to quantum metrology problems can be found in \cite{Rafal2015}, and a reader already familiar with these ideas can skip straight to the methodology in Section \ref{method} and our main results in Section \ref{results}.

\subsection{Uncertainty in single-parameter estimation}

Given an experiment where $\boldsymbol{n} = (n_1, n_2, ..., n_{\mu})$ are the outcomes of $\mu$ independent observations, an estimation function $g(\boldsymbol{n})$ can be constructed to estimate the unknown parameter $\theta$. The precision of this procedure is expressed with an error function $\epsilon\left[g(\boldsymbol{n}),\theta\right]$, and the uncertainty averaging over the different values the underlying parameter can take as well as the different measurement outcomes that can be obtained is defined as \cite{Jaynes2003}
\begin{align}
\bar{\epsilon} = \int d\boldsymbol{n}d\theta p(\boldsymbol{n},\theta)\epsilon\left[g(\boldsymbol{n}),\theta\right],
\label{err}
\end{align}
where $p(\boldsymbol{n},\theta)$ is the joint probability density function for the variables of the experiment. In addition, the product rule implies that $p(\boldsymbol{n},\theta) = p(\theta)p(\boldsymbol{n}|\theta)$. The function $p(\theta)$ is the prior probability density, and it encodes what is known about the parameter before the experiment is performed. This information can be given, for instance, by the results of previous experiments, and it will typically include the domain $a \leqslant \theta\leqslant b$ in which we can expect to find the parameter. The information about the outcomes of the actual experiment is encoded in the likelihood function $p(\boldsymbol{n}|\theta)$, and for a quantum system, the Born rule establishes that 
\begin{align}
p(\boldsymbol{n}|\theta) = \prod_{i=1}^{\mu}p(n_i|\theta) =\prod_{i=1}^{\mu} \mathrm{Tr}\left[E_{n_i}\rho(\theta)\right],
\label{likelihood}
\end{align}
where we have considered the following protocol: 

\begin{enumerate}
\item A probe state $\rho_0$ is prepared.
\item The parameter is encoded by means of some unitary interaction $U(\theta)$, producing the transformed state $\rho(\theta) = U(\theta)\rho_0 U^{\dagger}(\theta)$.
\item A positive-operator valued measure $E_{n_i}$ is used to model the measurement scheme.
\item The previous three steps are repeated $\mu$ times.
\end{enumerate} 

When the parameter to be estimated is periodic, as is the case for optical phase shifts, a periodic error function is the most suitable choice. The simplest option that satisfies the requirements of this symmetry is \cite{Rafal2015}
\begin{equation}
\epsilon\left[g(\boldsymbol{n}),\theta\right] = 4~ \mathrm{sin}^2\left[ \frac{g(\boldsymbol{n}) - \theta}{2} \right].
\label{sinerr}
\end{equation}
However, since $\mathrm{sin}^2(x) \approx x^2$ when $x$ is small, for a parameter domain less than one period Eq.~\ref{err} can be approximated as 
\begin{align}
\bar{\epsilon} \approx \bar{\epsilon}_{\mathrm{mse}} = \int d\boldsymbol{n}d\theta p(\boldsymbol{n},\theta)\left[g(\boldsymbol{n}) - \theta\right]^2,
\label{errwork}
\end{align}
which is the mean square error \footnote{In the literature this quantity is usually called \emph{average mean square error} to distinguish it from the analogous error used in non-Bayesian scenarios \cite{kay1993, Rafal2015}. Nevertheless, this distinction is not necessary in our work because we are considering a single measure of uncertainty plus its asymptotic approximation.}. The limitations of this approximation are discussed in Appendix~\ref{circular} for the specific scenarios considered in Section \ref{results}.

Assuming that the prior of the experiment is given and the encoding operator is known, the optimization of the metrology protocol is achieved by minimizing Eq.~\ref{errwork} with respect to the estimator, the measurement scheme and the probe state.

\subsection{Classical optimization: estimator and the asymptotic regime}\label{approx_classical}

If we look at Eq.~\ref{errwork} as a functional of $g(\boldsymbol{n})$, then the optimal estimator is determined classically by solving the variational problem \cite{Jaynes2003} 
\begin{align}
\delta \bar{\epsilon}_{\mathrm{mse}}\left[g(\boldsymbol{n}) \right] = \delta \int d\boldsymbol{n}~ \mathcal{L}\left[\boldsymbol{n},g(\boldsymbol{n})\right]= 0,
\end{align}
where $\mathcal{L}\left[\boldsymbol{n}, g(\boldsymbol{n})\right] = \int d\theta p(\boldsymbol{n},\theta) \left[g(\boldsymbol{n}) - \theta\right]^2$. As a result we have that
\begin{align}
g(\boldsymbol{n}) = \int d\theta p(\theta|\boldsymbol{n})\theta,
\label{optestimator}
\end{align}
where
\begin{align}
p(\theta|\boldsymbol{n}) = \frac{p(\theta)p(\boldsymbol{n}|\theta)}{\int d\theta p(\theta)p(\boldsymbol{n}|\theta)}
\label{bayes}
\end{align}
is the posterior density function defined by means of the Bayes theorem. Hence, Eq.~\ref{errwork} becomes
\begin{align}
\bar{\epsilon}_{\mathrm{mse}} = \int d\boldsymbol{n} p(\boldsymbol{n}) \epsilon(\boldsymbol{n}),
\label{erropt}
\end{align}
with $p(\boldsymbol{n}) = \int d\theta p(\theta)p(\boldsymbol{n}|\theta)$ and
\begin{equation}
\epsilon(\boldsymbol{n}) = \left\lbrace \int d\theta p(\theta|\boldsymbol{n}) \theta^2- \left[\int d\theta p(\theta|\boldsymbol{n}) \theta  \right]^2 \right\rbrace.
\label{errobayes}
\end{equation}
Note that Eq.~\ref{errobayes} is the variance of the parameter with respect to the posterior for the experimental data $\boldsymbol{n}$ \cite{Rafal2015}.

The calculation of Eq.~\ref{erropt} is still very challenging in general, and therefore it is important to identify further approximations that simplify the problem in practice. To accomplish that task, let us imagine a hypothetical scenario where the likelihood $p(\boldsymbol{n}|\theta)$ as a function of $\theta$ becomes narrower and concentrated around a maximum whose value is the unknown parameter $\theta'$ when $\mu \gg 1$. In addition, the prior knowledge is enough to identify a region of the parameter domain in which $\theta'$ can be found, although the experimental information dominates in this regime. In that case, the posterior function $p(\theta|\boldsymbol{n})$ can be approximated by the Gaussian density \cite{cox2000, Jaynes2003}
\begin{equation}
p(\theta|\boldsymbol{n}) \approx \sqrt{\frac{\mu F(\theta')}{2\pi}} \mathrm{exp} \left[ -\frac{\mu F(\theta')}{2}(\theta - \theta')^2 \right],
\label{gaussian}
\end{equation}
where
\begin{equation}
F(\theta) = \int dn p(n|\theta) \left\lbrace \frac{\partial \mathrm{log} \left[p(n|\theta)\right]}{\partial \theta}\right\rbrace^2
\label{fisherinfo}
\end{equation}
is the Fisher information and $n$ is the outcome for a single observation. Moreover, we further assume that the Fisher information does not depend on the parameter, so that $F(\theta) = F$ for all $\theta$. Thus we are able to approximate Eq.~\ref{erropt} as
\begin{equation}
\bar{\epsilon}_{\mathrm{mse}} \approx \frac{1}{\mu F}.
\label{crb}
\end{equation}

This result is known as \emph{Cram\'{e}r-Rao bound} in the context of local estimation theory \cite{kay1993, Rafal2015, paris2009}, although here we are using it as an approximation under certain circumstances to the Bayesian uncertainty defined by Eq.~\ref{err} and Eq.~\ref{sinerr}, and not as a proper bound. More concretely, Eq.~\ref{crb} holds when the number of observations $\mu$ is very large and the prior information is enough to localize the relevant domain. These properties define the asymptotic regime.

The details of this known heuristic argument are reviewed in Appendix~\ref{heuristic}. Furthermore, a more rigorous approach based on the theory of local asymptotic normality can be found in \cite{lecam1986,vaart1998}.

\subsection{Quantum optimization: measurement scheme and probe state}

According to Eq.~\ref{fisherinfo}, the Fisher information only depends on the likelihood function, which is constructed out of the measurement scheme and the transformed state. By maximising it over all the positive-operator value measures, it is possible to prove the inequality \cite{helstrom1967mmse, Helstrom1976, BraunsteinCaves1994, genoni2008}
\begin{equation}
F(\theta) \leqslant F_q(\theta) = \mathrm{Tr}\left[\rho(\theta)L(\theta)^2\right],
\label{quantumfisher}
\end{equation}
where $F_q(\theta)$ is the quantum Fisher information and the symmetric logarithmic derivative $L(\theta)$ satisfies
\begin{equation}
L(\theta)\rho(\theta) + \rho(\theta)L(\theta) = 2 \frac{\partial \rho(\theta)}{\partial \theta}.
\end{equation}
This bound is saturated if the measurement scheme is given by the projections onto the eigenstates of $L(\theta)$ \cite{BraunsteinCaves1994, genoni2008}.

Since the parameter is encoded with a unitary transformation, the quantum Fisher information will not depend on $\theta$ explicitly \cite{Rafal2015}. In that case, the saturation of Eq.~\ref{quantumfisher} implies that the approximation in Eq.~\ref{crb} becomes
\begin{equation}
\bar{\epsilon}_{\mathrm{mse}} \approx \bar{\epsilon}_{cr} = \frac{1}{\mu F_q},
\label{qcrb}
\end{equation}
which is known as \emph{quantum Cram\'{e}r-Rao bound} in the local approach \cite{Rafal2015, paris2009}. From this we can conclude that the asymptotic optimal precision is a function of $\rho(\theta)$ alone and that to find optimal probes in this regime we just need to maximize the quantum Fisher information.  

Nevertheless, from a physical perspective the number of observations is always limited by the available resources. In consequence, whenever two strategies are being compared in terms of the quantum Cram\'{e}r-Rao bound, in general it is also necessary to indicate how large $\mu$ needs to be such that Eq.~\ref{qcrb} is a good approximation. Moreover, if the likelihood reaches its maximum for several values of the parameter, then we need enough prior knowledge to select a single peak. The verification of the fulfilment of these crucial restrictions is not always done in the literature, a problem that can be overcome by using the framework of the next section.

\section{Methodology} \label{method}

The procedure described in Section~\ref{theory} does not specify the order of magnitude of $\mu$ nor the minimum prior knowledge that this strategy requires. Although the early proposal of \cite{braunstein_gaussian1992} answers to the former question by generalizing the likelihood equation in the local context and \cite{tsang2012} catches the influence of the prior probability to some extent, there is not a method that takes into account the combined action of these restrictions simultaneously and exactly in practical scenarios. This motivates the search of a more general approach. A solution to this problem is provided by combining different known Bayesian techniques into a pragmatic methodology.

\subsection{Experimental configuration and prior knowledge}\label{experiment_prior}

Let us consider that we arrange an experiment such that a system described by $\rho(\theta)$ is measured with a scheme that is optimal with respect to the quantum Cram\'{e}r-Rao bound. This configuration is then summarized with $p(\boldsymbol{n}|\theta)$ through Eq.~\ref{likelihood}.

On the other hand, in Section \ref{theory} we discussed that the likelihood function needs to be concentrated around its highest peak in order to be able to use the approximation in Eq.~\ref{crb} (see also the construction reviewed in Appendix~\ref{heuristic}). This local behaviour implies that, for a given scheme, the width of the parameter domain must be such that the solution to the problem $\partial p(\boldsymbol{n}|\theta)/\partial \theta = 0$ includes an asymptotically unique absolute maximum. Hence, we introduce the quantity $W_{\mathrm{int}}$, which we call \emph{intrinsic width}, and we define it as the width that fulfils the above criterion on average. Notice that if $W_0 > W_{\mathrm{int}}$, where $W_0$ is the initial width of our scheme, then the experiment cannot distinguish between two or more equally likely values, and the mean square error tends to a constant when $\mu \gg 1$.

In practice, the prior information is determined by the experimental configuration under consideration. We will see that different states are associated to a different $W_{\mathrm{int}}$; consequently, only those states with a value for $W_{\mathrm{int}}$ that is greater than or equal to the width imposed by the experiment  would be useful in a real scenario. 

For optical phases, and assuming that the only information known a priori about the parameter includes the length of the relevant domain, a flat prior is a reasonable choice, since it does not modify the information of the likelihood in the region where it becomes narrower. In addition, it simplifies the calculations. Therefore, we will consider that this probability distribution is the uninformative \emph{intrinsic prior} of our particular strategy, and we will use it for our analysis \footnote{Although we have chosen a semi-uninformative scenario, the methodology proposed in this work can be also applied to more realistic cases. For instance, we could imagine that our experiment was previously carried by a different team and that we have a summary of their findings encoded in the probability density $p(\theta)$.}.

To find $W_{\mathrm{int}}$ we can plot the posterior probability $p(\theta| \boldsymbol{n})$ as a function of $\theta$ directly, since its relative extremes coincide with those of the likelihood when the prior is flat. This procedure depends on the simulation of several random outcomes $\boldsymbol{n}$ for different values of the parameter, and thus the solution is necessarily probabilistic. However, this is enough for our purposes because our analysis only requires that this is satisfied in the asymptotic regime, where $\mu$ is large.

\subsection{Numerical strategy}

We now have all the pieces that are necessary to calculate Eq.~\ref{erropt} exactly, which is the next step of our strategy. Since this integral has $(\mu+1)$ dimensions and we are interested in studying its behaviour as $\mu$ increases, in general we can only compute it numerically. While this is a purely numerical problem that arises in the Bayesian literature \cite{Jaynes2003, Rafal2015} and can be treated with well known numerical techniques \cite{mathematics2004, numerics2014matlab}, we believe that giving an explicit scheme of calculation in terms of physical arguments as part of the methodology offers conceptual clarity and insight. In particular, we have followed a three-step method:

\begin{enumerate}
\item If a collection of $\mu$ experimental outcomes $\boldsymbol{n}$ was originated from the unknown parameter $\theta'$, and assuming the knowledge of $p(\boldsymbol{n}|\theta)$ and $W_{\mathrm{int}}$ previously discussed, then the error of the estimation based on that particular experiment will be given by Eq.~\ref{errobayes}, that is, by the variance of the posterior probability $p(\theta|\boldsymbol{n})$. Moreover, this uncertainty is understood in \cite{PaulProctor2016} as the error that arises from gathering and processing data in a real experiment. The integral that defines this quantity can be calculated with a standard deterministic method after the simulation of $\boldsymbol{n}$ for a given $\theta'$, which implies that Eq.~\ref{errobayes} depends on $\theta'$ through the values of the outcomes.

\item According to Eq.~\ref{errobayes}, different uncertainties $\epsilon(\boldsymbol{n})$ can be associated to the estimation depending on the particular values $\boldsymbol{n}$. Therefore, if our aim is to simulate experiments whose performance is optimal on average, we need to calculate the average of the errors for all the possible experimental outcomes associated with $\theta'$ weighted by their likelihood, i.e.,
\begin{equation}
\epsilon(\theta') = \int d\boldsymbol{n}p(\boldsymbol{n}|\theta')\epsilon(\boldsymbol{n}).
\label{simulation}
\end{equation}
This is precisely what is done in \cite{PaulProctor2016}. The multidimensional integral in Eq.~\ref{simulation} can be solved using Monte Carlo techniques \cite{mathematics2004, numerics2014matlab}.

\item The previous quantity still depends on $\theta'$, which is not known. However, by taking the average
\begin{eqnarray}
&\int & d\theta' p(\theta')\epsilon(\theta')=\bar{\epsilon}_{\mathrm{mse}}
\label{finalmse}
\end{eqnarray}
weighted over our prior knowledge of $\theta'$ we finally obtain the mean square error, which is independent of the values of both the parameter and the outcomes. Following the previous discussion, $\bar{\epsilon}_{\mathrm{mse}}$ represents the uncertainty on average about the knowledge that we can acquire in principle with the experimental configuration that is being studied, and as such it is the suitable figure of merit to design experiments from theoretical considerations. The integral over $\theta'$ can be calculated by a deterministic numerical method once $\epsilon(\theta')$ is known for different values of $\theta'$ from the second step.
\end{enumerate}

Although there are other ways of implementing this calculation \footnote{This is because we can rearrange the integrals of the mean square error depending on how we split the joint probability $p(\boldsymbol{n},\theta)$, which can be expressed either as $p(\theta)p(\boldsymbol{n}|\theta)$ or $p(\boldsymbol{n})p(\theta|\boldsymbol{n})$. In spite of the fact that they are theoretically equivalent, changing the order in which we integrate the variables changes the numerical performance.}, the reason to choose the strategy described above is twofold. Firstly, it offers a clear physical motivation for the use of the measure of uncertainty defined in Eq.~\ref{err} as the figure of merit. Secondly, its numerical implementation is relatively straightforward, and it has turned out to be robust against small variations of the numerical parameters for a reasonable number of iterations. 

\subsection{Classical approximation threshold}

Our final goal is to quantify the deviation of the quantum Cram\'{e}r-Rao bound as a function of the number of observations. Once we know the exact value of Eq.~\ref{erropt} for our particular scheme, a simple way of achieving this is to introduce the relative error
\begin{equation}
\frac{\varepsilon_{\tau} \%}{100 \%} = \frac{\abs{\bar{\epsilon}_{\mathrm{mse}} - \bar{\epsilon}_{cr}}}{\bar{\epsilon}_{\mathrm{mse}}},
\label{saturation}
\end{equation}
for $\bar{\epsilon}_{\mathrm{mse}} \neq 0$. This will give us the minimum number of observations $\mu_{\tau}$ that is needed such that the approximation in Eq.~\ref{qcrb} is valid for a given threshold $\varepsilon_{\tau}$, which should be chosen according to the requirements of the specific experimental configuration that is being analysed.

\begin{figure*}[t]
\centering
\includegraphics[trim={0.65cm 0.1cm 1.5cm 0.5cm},clip,width=9.1cm]{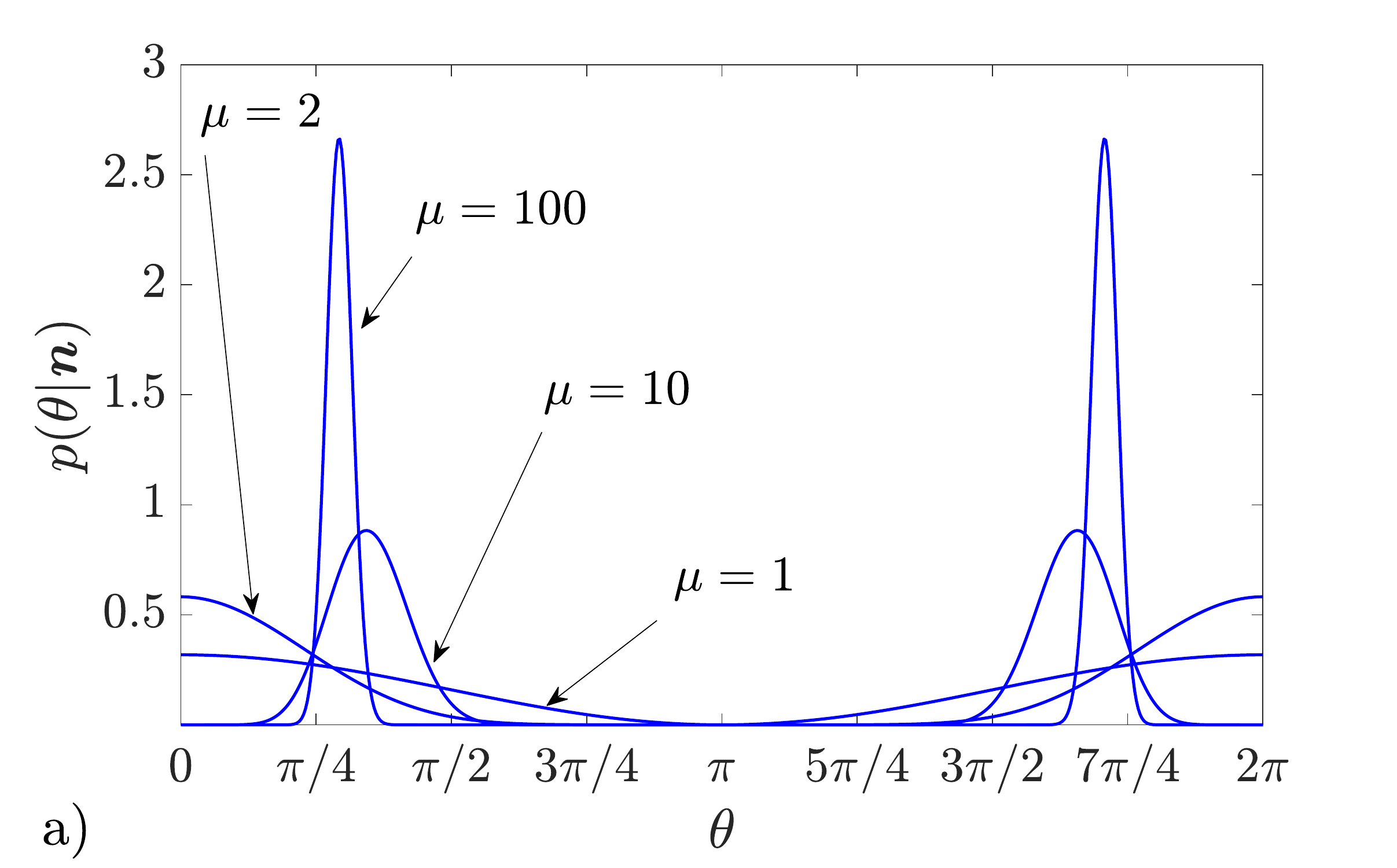}\includegraphics[trim={0.65cm 0.1cm 1.5cm 0.5cm},clip,width=9.1cm]{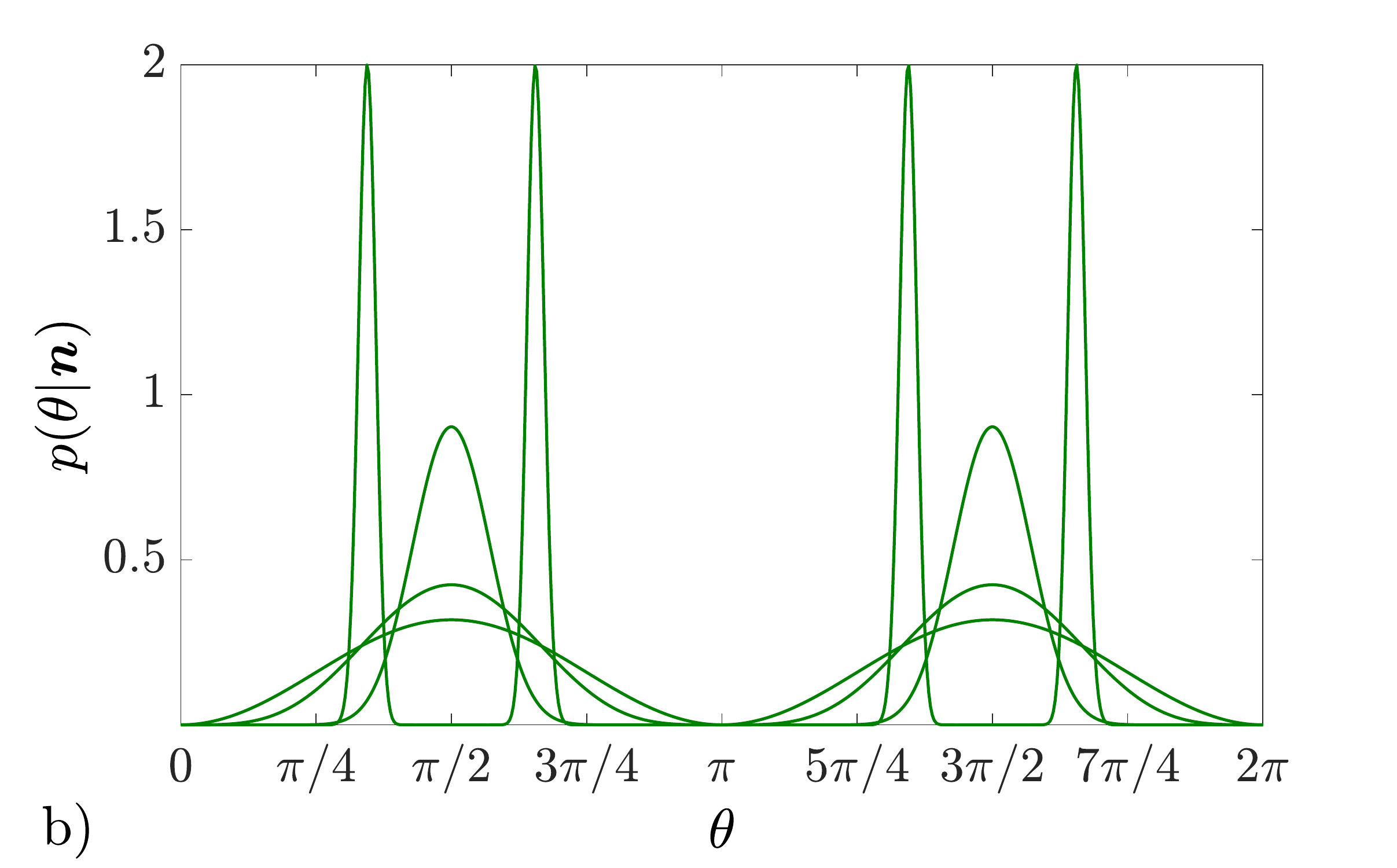}
\includegraphics[trim={0.65cm 0.1cm 1.5cm 0.5cm},clip,width=9.1cm]{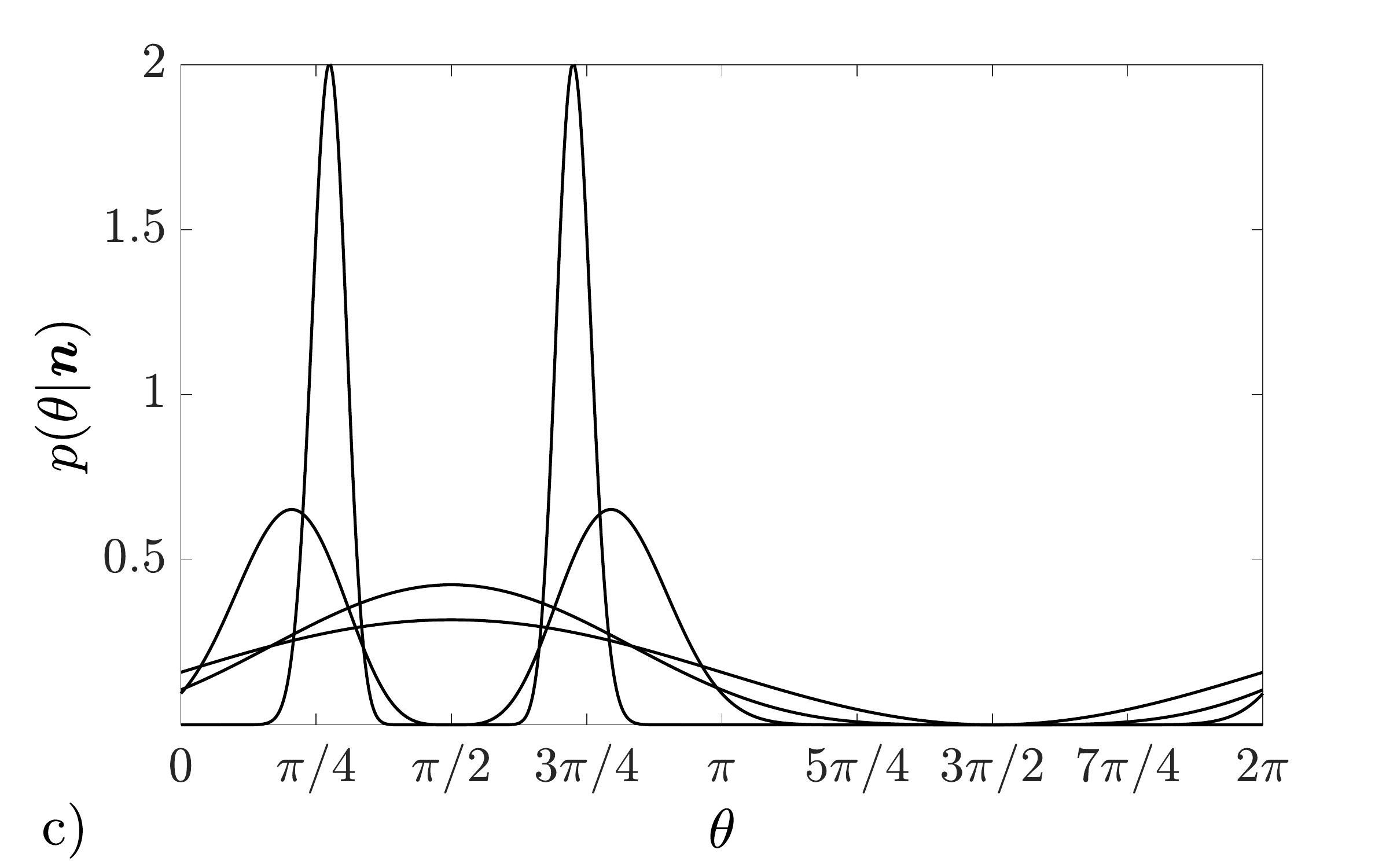}\includegraphics[trim={0.65cm 0.1cm 1.5cm 0.5cm},clip,width=9.1cm]{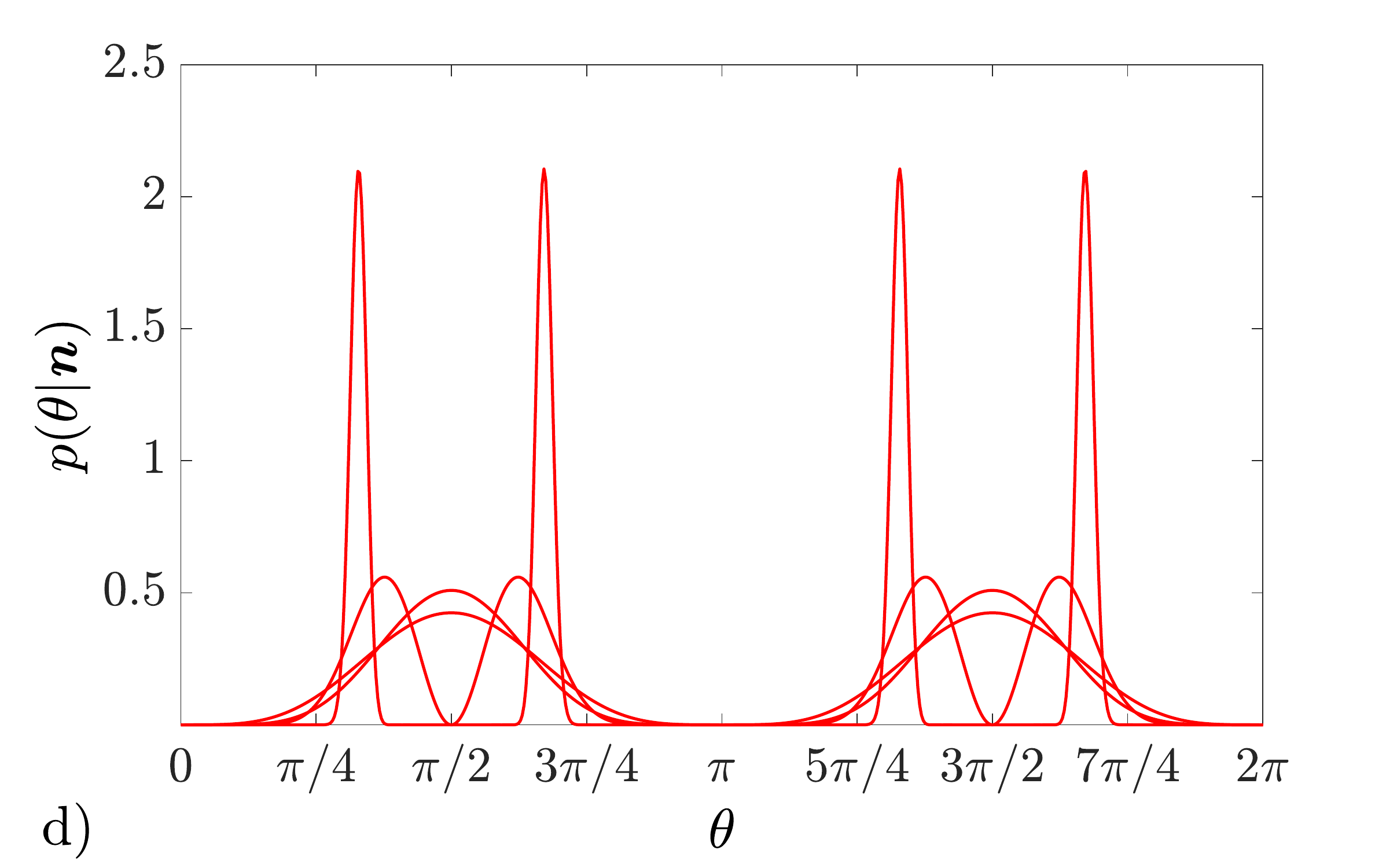}
	\caption{Posterior density functions for random simulations of 1, 2, 10 and 100 observations, a flat prior and a photon-counting measurement implemented after the action of a 50:50 beam splitter. The initial probes are: a) coherent state with $\bar{n} = 2$, b) NOON state with $\bar{n} = 2$, c) NOON state with $\bar{n} = 1$, and d) twin squeezed vacuum with $\bar{n} = 2$. We draw attention to the fact that these configurations cannot distinguish a unique value  when the initial prior is set to $W_0 = 2\pi$, even if we are in the asymptotic regime with $\mu \gg 1$.}
\label{prior}
\end{figure*}

\subsection{Bayesian quantum bounds}

A different approach that can also identify the situations in which the Cram\'{e}r-Rao bound fails is based on deriving alternative quantum bounds that are valid for all $\mu$. This idea was precisely explored in \cite{tsang2012,tsang2016}, where the two main families of classical Bayesian bounds \cite{bayesbounds2007} were extended to the quantum regime. According to their results, the quantum Ziv-Zakai bound for a flat prior between $a = 0$ and $b = W_0$ is \cite{tsang2012}
\begin{equation}
\bar{\epsilon}_{\mathrm{mse}} \geqslant \frac{1}{2} \int d\theta \theta \left(1 - \frac{\theta}{W_0} \right) \left[1 - \sqrt{1 - \abs{f(\theta)}^{2\mu}} \right],
\label{qzzb}
\end{equation}
where $f(\theta) = \bra{\psi_0}\ket{\psi(\theta)}$, $\ket{\psi_0}$ is a pure state and $\ket{\psi(\theta)}$ encodes the parameter with a unitary transformation. In addition, the quantum Weiss-Weinstein bound establishes that \cite{tsang2016}
\begin{equation}
\bar{\epsilon}_{\mathrm{mse}} \geqslant \sup_{\theta} \frac{\theta^2 \left(1-\frac{\theta}{W_0}\right)^2 \abs{f(\theta)}^{4\mu}/2}{\abs{f(\theta)}^{2\mu} - \left(1-\frac{2\theta}{W_0}\right)\mathrm{Re}\left\lbrace \left[f(\theta)^{2} {f(2\theta)}^{*}\right ]^\mu \right\rbrace}.
\label{qwwb}
\end{equation}

There also exists a Bayesian version of the Cram\'{e}r-Rao bound based on the van Trees inequality \cite{gill1995}. Unfortunately, its derivation requires that the prior satisfies the boundary conditions $p(a) \rightarrow 0$ and $p(b) \rightarrow 0$, and this excludes the case of the flat prior between $a$ and $b$.

In spite of the utility of this method, the key advantage of using the direct calculation of the mean square error instead is that then we are evaluating the validity of the Cram\'{e}r-Rao bound exactly. Nevertheless, we will still make use of these bounds as a consistency test for the numerical evaluation of Eq.~\ref{erropt}.

\section{Results and discussion} \label{results}

The methodology that we have described is general enough to accommodate a wide range of estimation problems, and in this section we explore its application to phase estimation in optical interferometry \cite{Rafal2015,yurke1986}. These results constitute the main contribution of this work.

Let us assume that we are working in the number basis of a two-path interferometer, and that the parameter $\theta$ is encoded as a difference of phase shifts by means of the unitary transformation $
U(\theta) = \mathrm{exp}[-i(a_1^{\dagger}a_1 - a_2^{\dagger}a_2)\theta/2]$, where $a_i,a_i^{\dagger}$ are the creation and annihilation operators for the modes $i = 1,2$. Here we focus on a collection of states that together represent the common techniques currently used in quantum metrology \cite{Rafal2015, ShaotaQuesada2015, PaulProctor2016, jarzyna2012}. Concretely, we consider:
\begin{enumerate}
\item Coherent states
\begin{eqnarray}
\ket{\psi_0} &=& U_{\mathrm{BS}} D(\alpha)\otimes\mathbb{I} \ket{0, 0} 
\nonumber \\
&=& |\alpha/\sqrt{2},-i\alpha/\sqrt{2}\rangle,
\end{eqnarray}
where $U_{\mathrm{BS}} = \mathrm{exp}[-i(a_1^{\dagger}a_2 + a_2^{\dagger}a_1)\pi/4]$ is a 50:50 beam splitter and $D(\alpha) = \mathrm{exp}(\alpha a_1^{\dagger} - \alpha^{*}a_1)$ is the displacement operator.
\item NOON states
\begin{eqnarray}
\ket{\psi_0} &=& \frac{1}{\sqrt{2}}(|N, 0\rangle + |0, N\rangle).
\label{noon}
\end{eqnarray}
\item Twin squeezed vacuum
\begin{align}
\ket{\psi_0} = S_1(r) S_2(r)|0, 0\rangle = |r,r\rangle,
\end{align}
where $S_i(r) = \mathrm{exp}\{[r^{*}a_i^2 - r (a_i^{\dagger})^2]/2\}$, for $i = 1, 2$, are squeezing operators. 
\item Squeezed entangled states
\begin{eqnarray}
\ket{\psi_0} &=& \mathcal{N} (|r,0\rangle +|0,r\rangle),
\end{eqnarray}
where $\mathcal{N} = \left[2 + 2/\mathrm{cosh}(\abs{r})\right]^{-1/2}$.
\end{enumerate}
Since coherent states present no quantum correlations, their precision is asymptotically given by the standard quantum limit. Contrarily, NOON states have inter-mode and intra-mode correlations and can achieve the Heisenberg limit, although the twin squeezed vacuum also achieves a Heisenberg scaling having intra-mode correlations only \cite{Rafal2015, ShaotaQuesada2015}. Finally, the squeezed entangled states, which have both types of correlations, constitute a precision improvement over the previous states \cite{PaulProctor2016}. Note that we have selected pure states for the sake of simplicity, but our methods would be also applicable to mixed states.

A common property of these configurations is that they belong to the family of path-symmetric states introduced in \cite{HofmannHolger2009}. Therefore, their classical Fisher information will reach the bound imposed in Eq.~\ref{quantumfisher} by its quantum counterpart if we implement a photon-counting measurement after the action of a 50:50 beam splitter. This implies that any discrepancy between Eq.~\ref{erropt} and Eq.~\ref{qcrb} must necessarily come from the approximation that we discussed in Section~\ref{approx_classical}.  

\begin{figure*}[t]
\centering
\includegraphics[trim={0.4cm 0cm 0.8cm 0.4cm},clip,width=9.4cm]{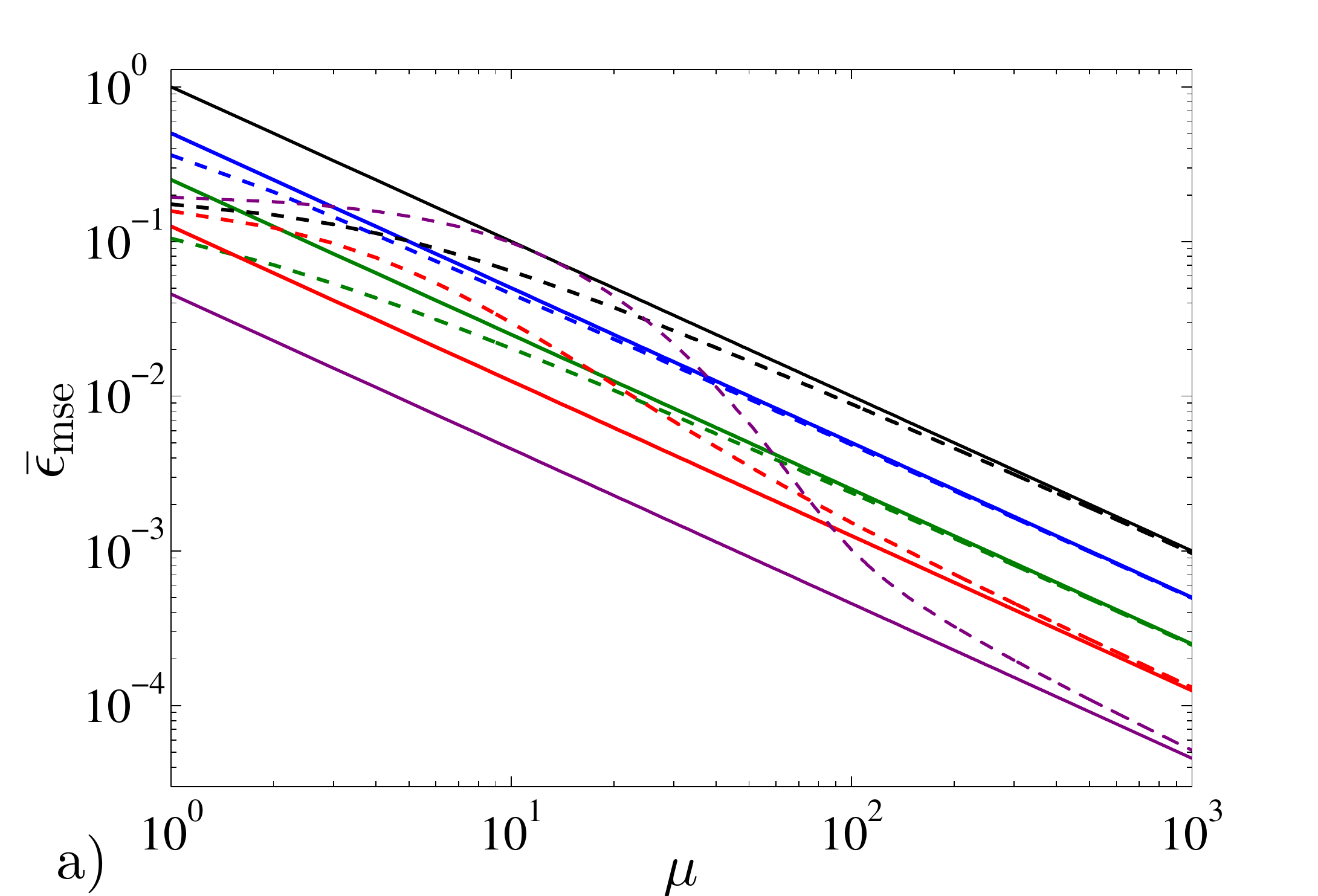}\includegraphics[trim={1.2cm 0cm 0 0.4cm},clip,width=9.4cm]{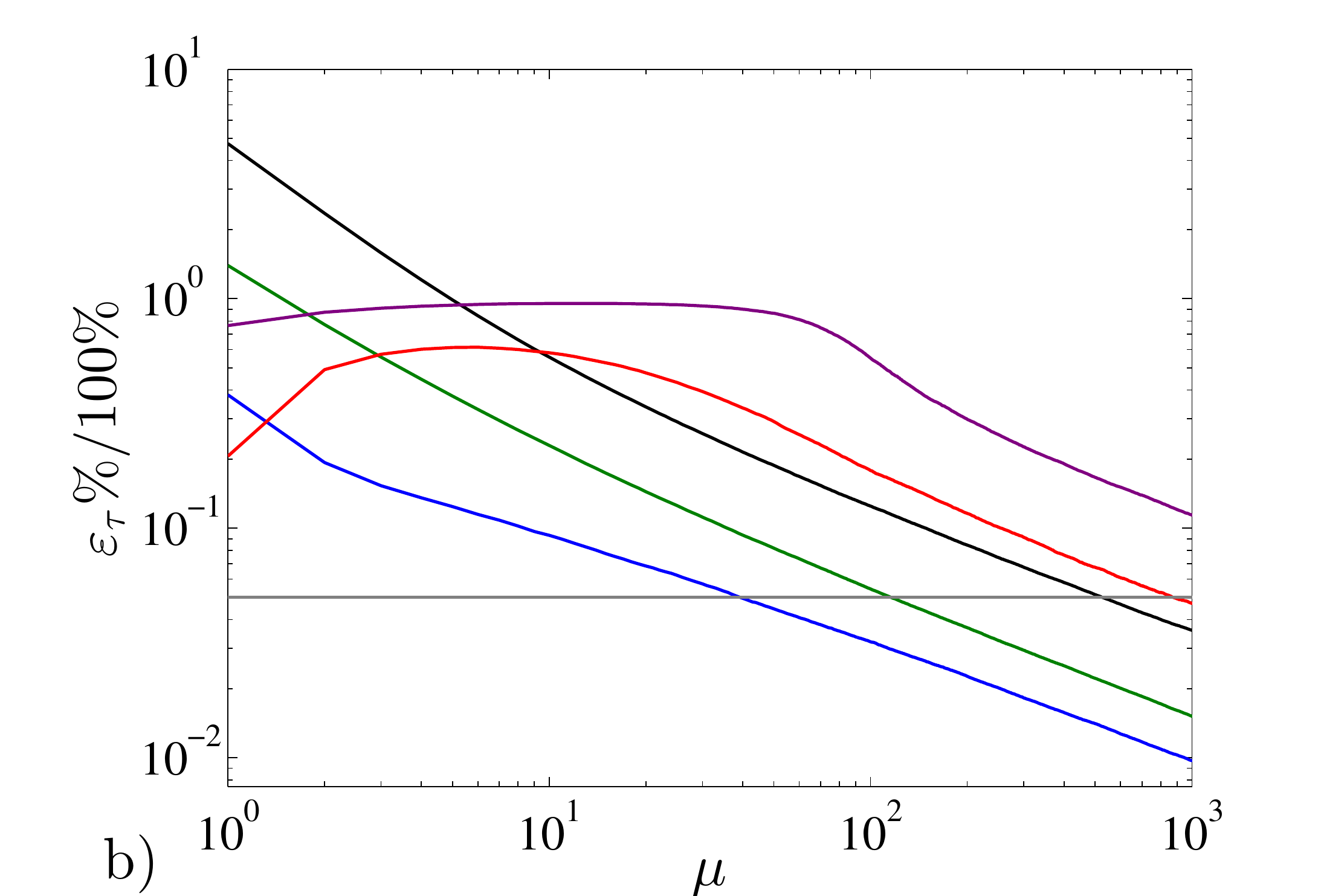} 
\includegraphics[trim={0.4cm 0cm 0.8cm 0.4cm},clip,width=9.4cm]{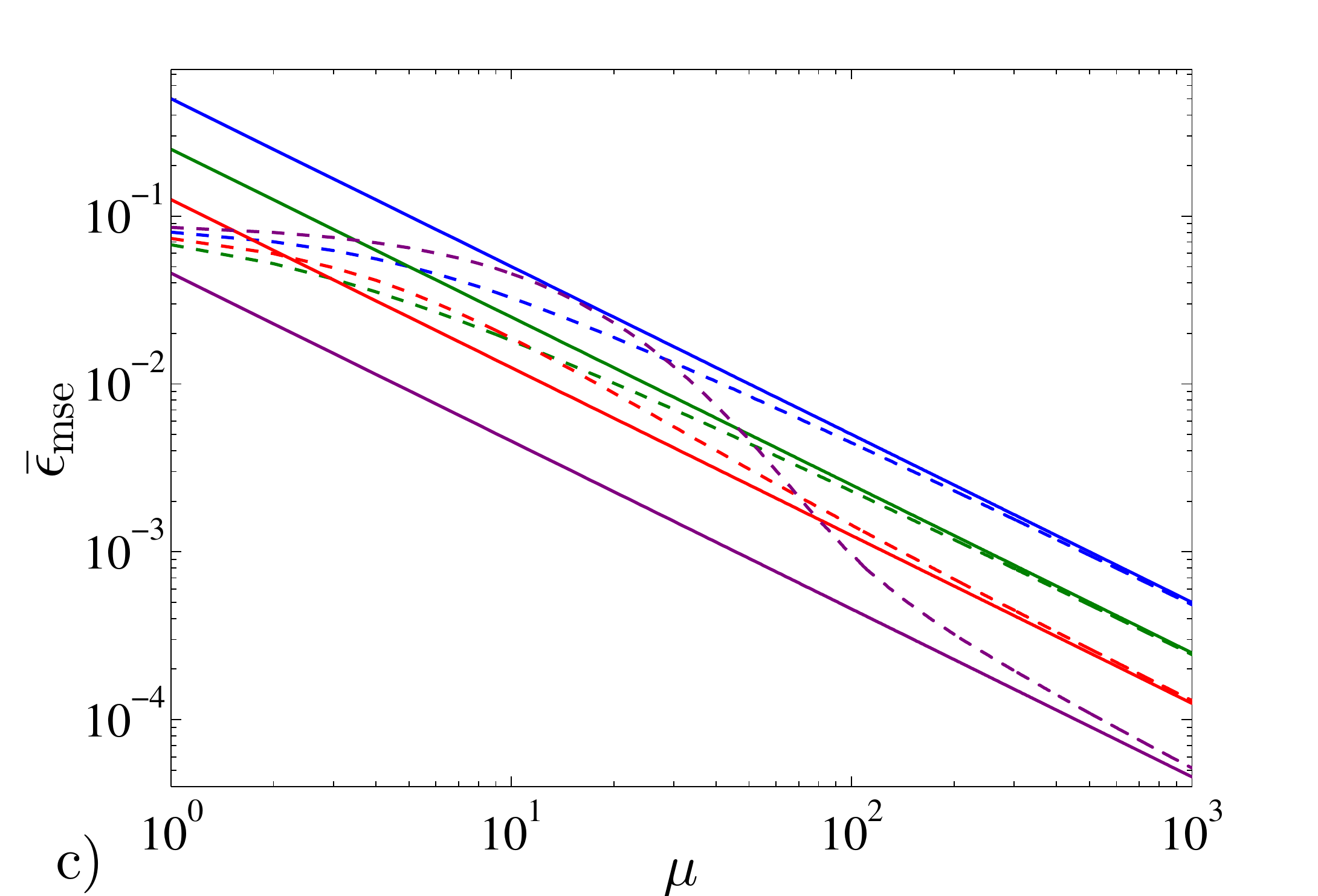}\includegraphics[trim={1.2cm 0cm 0 0.4cm},clip,width=9.4cm]{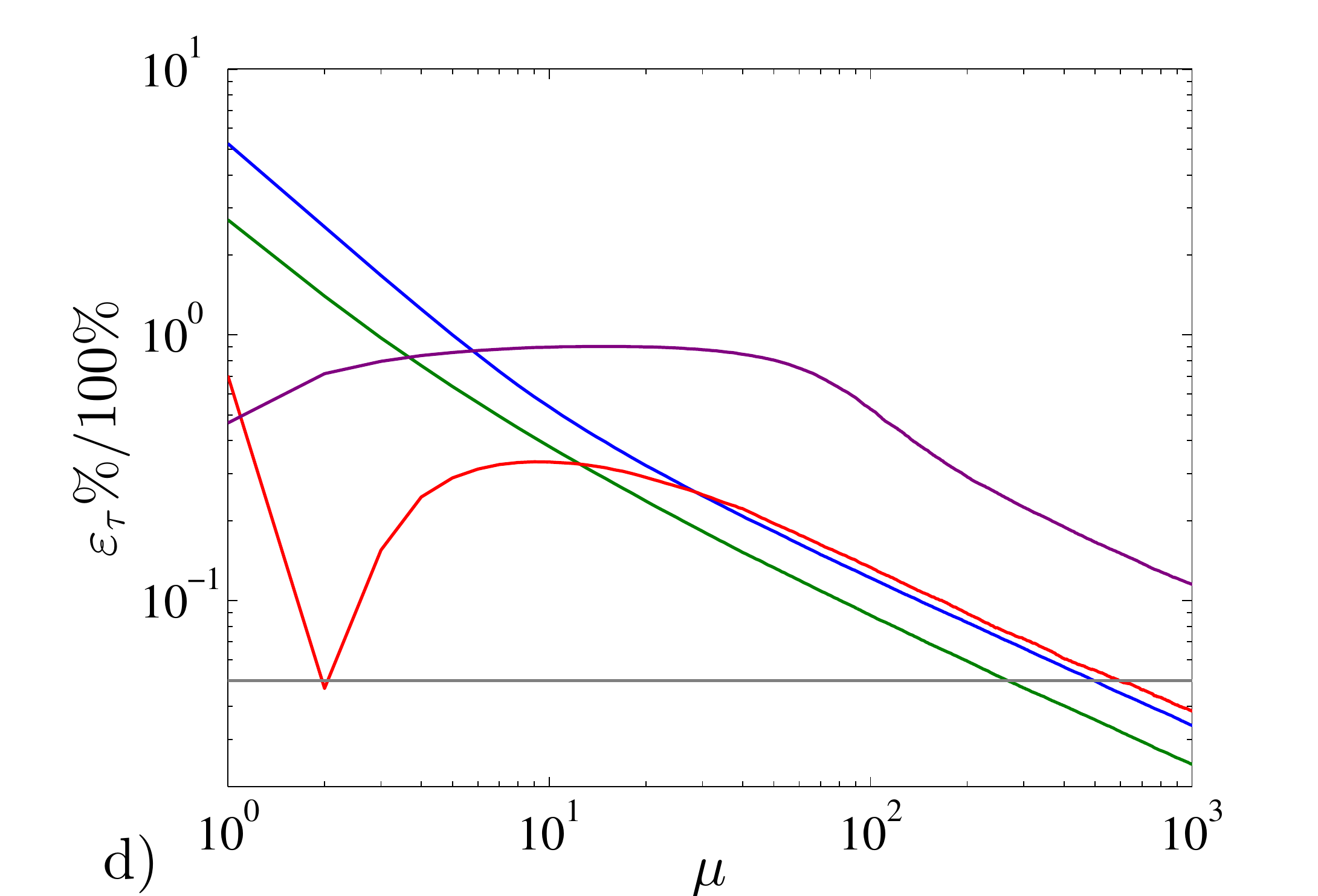}
\caption{a) Quantum Cram\'{e}r-Rao bound (solid line) and optimal mean square error (dashed line) for a coherent state with $\bar{n} = 2$ and $W_{\mathrm{int}} = \pi$ (blue line), a NOON state with $\bar{n} = 2$ and $W_{\mathrm{int}} = \pi/2$ (green line), a NOON state with $\bar{n} = 1$ and $W_{\mathrm{int}} = \pi/2$ (black line), a twin squeezed vacuum with $\bar{n} = 2$ and $W_{\mathrm{int}} = \pi/2$ (red line), and a squeezed entangled state with $\bar{n} = 2$ and $W_{\mathrm{int}} = \pi/2$ (purple line), where $\bar{n}$ is the mean number of quanta per observation and $W_{\mathrm{int}}$ is the intrinsic width; b) relative error defined by Eq.~\ref{saturation} with a threshold $\varepsilon_\tau = 5$ (grey line) for the states considered in Figure~\ref{mainresult}.a; c) repetition of the calculation performed in Figure~\ref{mainresult}.a with a common prior width $W_0 = \pi/3$ and the same values for $\bar{n}$; and d) relative error for the states considered in Figure~\ref{mainresult}.c. The consequences of these results are explored in the main text.}
\label{mainresult}
\end{figure*}

The first step to apply our numerical strategy is to identify the intrinsic width $W_{\mathrm{int}}$ of each state for a given mean number of particles per probe $\bar{n}$. Some of the random simulations that are required to achieve that goal are shown in Figure~\ref{prior}, which allow us to deduce the size of the maximum width by direct examination \footnote{An alternative way of determining $W_{\mathrm{int}}$ is to study the symmetries of the likelihood $p(n|\theta)$ as a function of $\theta$ for $\mu=1$. In our case, the values extracted from Figure~\ref{prior} can be explained by combining the periodicity of the phase and a reflection within each period. However, our method is useful to find this information even if the analytical expression for $p(n|\theta)$ is not available, which is sometimes the situation for more complicated states.}. For a twin squeezed vacuum and a squeezed entangled state we have found that $W_{\mathrm{int}} = \pi/2$, while coherent states have $W_{\mathrm{int}} = \pi$. The latter value was also determined by a different method in \cite{kolodynski2014}. Note that those results hold for any $\bar{n}$. On the contrary, with NOON states we have that $W_{\mathrm{int}} = \pi/\bar{n}$ or $W_{\mathrm{int}} = \pi/(2\bar{n})$ depending on whether the value for $N$ in Eq.~\ref{noon} is even or odd. It can be observed that none of the states allows us to uniquely identify the relative phase shift when we have no information about its possible values, that is, if $W_0= 2\pi$. Moreover, the NOON states present an intrinsic width smaller than $2\pi/\bar{n}$, which is their natural periodicity. We conclude then that the scheme that we are employing introduces some limitations to the estimation of the parameter, in spite of the fact that the measurement is optimal according to the quantum Cram\'{e}r-Rao bound criterion.

Once $W_{\mathrm{int}}$ is known, we calculate Eq.~\ref{erropt}, Eq.~\ref{qcrb} and Eq.~\ref{saturation} with the uniform prior 
\begin{align}
p(\theta) = 1/W_{\mathrm{int}},~\mathrm{for}~ \theta \in [0,W_{\mathrm{int}}],
\end{align}
and $p(\theta) = 0$ otherwise. The results are shown in Figure~\ref{mainresult}.a and Figure~\ref{mainresult}.b, where we have assumed that the experiment can only be repeated $\mu = 10^3$ times as an extra constraint. For this number of observations, the mean square error of coherent, NOON and twin squeezed vacuum states is close enough to the result predicted by the quantum Cram\'{e}r-Rao bound. In particular, their relative error is smaller than the selected threshold $\varepsilon_\tau = 5$. However, the minimum number of observations that are needed in order to reach that threshold is different for different states, and the squeezed entangled state does not even reach it in the regime that we are studying. This state-dependent phenomenon, whose concrete values are indicated in Table~I, has important consequences. 

If we consider first the comparison between a NOON state and a twin squeezed vacuum with $\bar{n} = 2$, $W_{\mathrm{int}} = \pi/2$, we can see that the latter is a better choice according to the Fisher information, but its error is higher for $\mu < 20$. Even if we focus on the results of the asymptotic regime, the twin squeezed vacuum requires $\mu \sim 10^3$ observations to achieve it, while the NOON state only needs $\mu \sim 10^2$. Thus a state whose Fisher information is maximum with respect to other probes can still produce a larger error if the experiment is operating outside of the asymptotic regime. Moreover, although it was shown that only the intra-mode correlations are crucial to surpass the standard quantum limit in the regime where the Fisher approach is valid \cite{ShaotaQuesada2015, sahota2016, proctor2017networked}, this comparison between a NOON state, which includes both types of correlations, and a twin squeezed vacuum, that has intra-mode correlations only, suggests that the role of quantum correlations in metrology should be revisited for the non-asymptotic regime.

On the other hand, a coherent state with $\bar{n} = 2$, $W_{\mathrm{int}} = \pi$ is less precise than a NOON state with $\bar{n} = 1$, $W_{\mathrm{int}} = \pi/2$ when $\mu \sim 1$. This implies that there is a region in which a probe with fewer resources can still beat a scheme with more photons if the prior knowledge of the former is higher. By combining these observations with those extracted from the previous probes we conclude that the Cram\'{e}r-Rao bound can both overestimate and underestimate the precision outside of its regime of validity. It is particularly relevant to draw attention to the latter case, since the fact that NOON and coherent states display a mean square error which is lower than their respective Cram\'{e}r-Rao bounds for low values of $\mu$ demonstrates that the unbiased estimators of the local theory are not always optimal \footnote{An estimator is called \emph{unbiased} in the local approach when $\int d\boldsymbol{n} p(\boldsymbol{n}|\theta) g(\boldsymbol{n}) = \theta$ is satisfied \cite{Rafal2015}. This technical condition is not usually crucial in Bayesian scenarios, and the estimators that we have calculated do not satisfy it in the non-asymptotic regime.}.

The analysis of the squeezed entangled state provides further details of the properties of the non-asymptotic regime. In particular, its performance is worse than all the previous cases for $\mu \sim 10$, and it only becomes the best choice when the number of repetitions is greater than $\mu \sim 10^2$. Surprisingly, this result is showing that while states with an indefinite number of photons can do better than the optimal choice for a finite number of quanta, NOON states have the best absolute precision among the cases that we have studied if the number of observations is less than $\mu \sim 10$.

To have a fairer comparison, we have also repeated the calculation with a common width $W_0 = \pi/3$ and $\bar{n} = 2$. Figure~\ref{mainresult}.c and Figure~\ref{mainresult}.d show that, while the numerical values are slightly different, the qualitative conclusions are the same. Nonetheless, there is an important difference given that the prior knowledge is now higher. For the NOON and coherent states, $\mu_{\tau}$ has increased with respect to the previous calculation, since the starting difference between the mean square error and the bound is now greater. On the other hand, for the twin squeezed vacuum there is a point where now the mean square error crosses the Cram\'{e}r-Rao bound before a stable saturation is reached. This happens because for $W_0 = W_{\mathrm{int}}$ the mean square error approached the bound from above, while for $W_0 = \pi/3$ the error begins below the bound and then crosses it to achieve the asymptotic regime from above. This suggests that if we keep increasing our prior information and we make the width of the parameter domain very small, then the number of observations needed to approach the Cram\'{e}r-Rao bound will grow. 

It is possible to formalize the previous phenomenon and derive an intuitive and informative relation that detects states that are not well-behaved. Firstly, we note that the uncertainty of an estimation that is made before we perform the experiment is represented by the variance of the prior probability 
\begin{equation}
(\bar{\epsilon}_{\mathrm{mse}})|_{\mu = 0} = \Delta \theta^2_p = \int d\theta p(\theta) \theta^2 - \left[\int d\theta p(\theta) \theta \right]^2,
\end{equation}
which is ${W_0}^2/12$ for a flat distribution of width $W_0$. On the other hand, we know that the precision is given by the Fisher information when $\mu \gg 1$; consequently, an estimation protocol is only worthwhile when
\begin{equation}
\Delta \theta^2_p(\rho) > \frac{1}{\mu(\rho) F_q(\rho)}
\label{criterion}
\end{equation}
is asymptotically satisfied, where we have made explicit the dependence on the state to indicate that the values of $\mu$ and $\Delta \theta^2_p$ guarantee that the Cram\'{e}r-Rao regime can be reached. If Eq.~\ref{criterion} were not fulfilled, then the experiment would not be telling us more than what we already knew. By reorganizing the terms we finally arrive to
\begin{equation}
\mu(\rho) > \frac{1}{\Delta \theta^2_p(\rho) F_q(\rho)},
\label{fundamental}
\end{equation}
which is a constraint based on practical requirements.

\begin{table} [t]
\begin{tabular}{|l|c|c|c|c|}
\hline
Probe state & $\bar{n}$ & $W_{\mathrm{int}}$ & $\mu_{\tau} (W_{\mathrm{int}})$ & $\mu_{\tau} (W_0=\pi/3)$\\
\hline
\hline
$|\alpha/\sqrt{2},-i\alpha/\sqrt{2}\rangle$ & $2$ & $\pi$ & $3.9\cdot 10$ & $4.97\cdot 10^2$\\
NOON state (even $N$) & $2$ & $\pi/2$ & $1.15\cdot 10^2$ &$2.67\cdot 10^2$\\
NOON state (odd $N$)& $1$ & $\pi/2$ & $5.26\cdot 10^2$ & - \\
$S_1(r)S_2(r)\ket{0,0}$ & $2$ & $\pi/2$ & $8.74\cdot 10^2$ & $5.95\cdot 10^2$\\
$\mathcal{N}(\ket{r,0}+\ket{0,r})$ & $2$ & $\pi/2$ & $>10^3$ &$ >10^3$\\
\hline
\end{tabular}
\caption{Numerical values of $W_{\mathrm{int}}$ and $\mu_{\tau}$ obtained in Figure~\ref{prior} and Figure~\ref{mainresult}, respectively,  for an asymptotically optimal strategy and a threshold $\varepsilon_\tau = 5$. The representation of the posterior probability $p(\theta|\boldsymbol{n})$ for the squeezed entangled state that provides the value of its intrinsic width was very similar to that of the twin squeezed vacuum, and therefore it has been omitted in Figure~\ref{prior} for brevity. In addition, note that we have chosen $\bar{n} = 2$ for most of our schemes in order to detect a significant improvement over the standard quantum limit.}
\label{table_summary}
\end{table}

\begin{figure*}
\centering
\includegraphics[trim={0.4cm 0cm 0.8cm 0.4cm},clip,width=9.2cm]{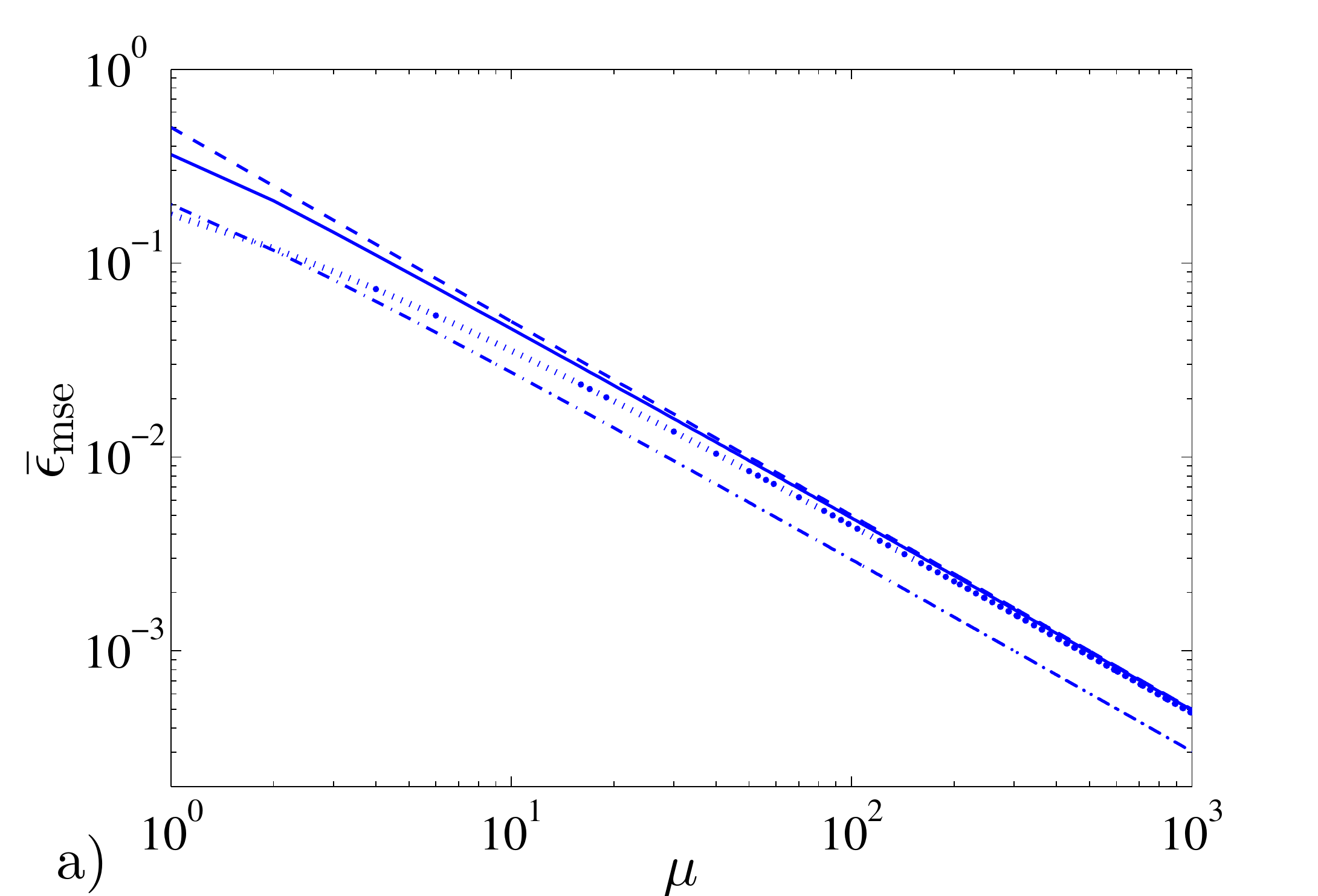}\includegraphics[trim={0.4cm 0cm 0.8cm 0.4cm},clip,width=9.2cm]{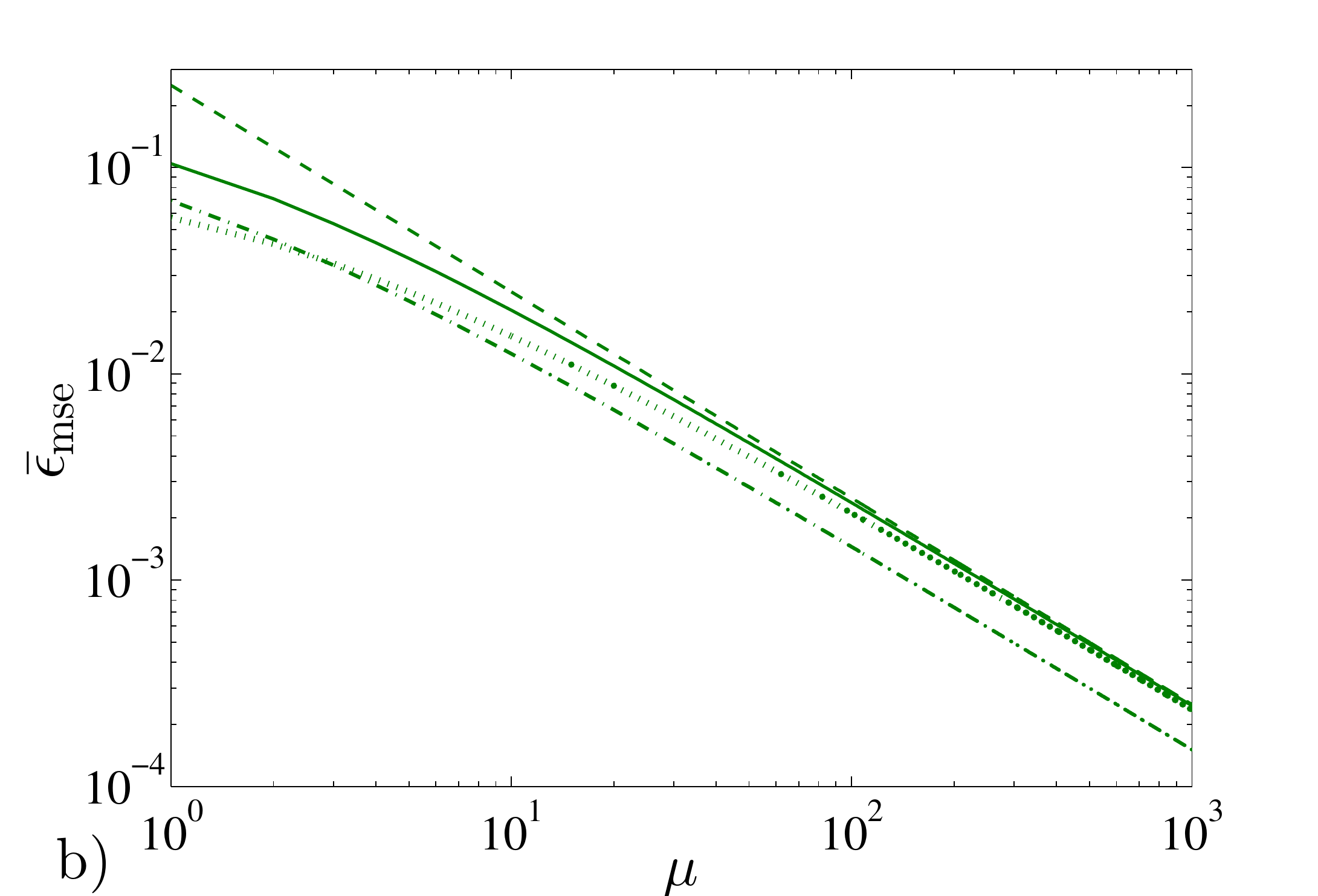}
\includegraphics[trim={0.4cm 0cm 0.8cm 0.4cm},clip,width=9.2cm]{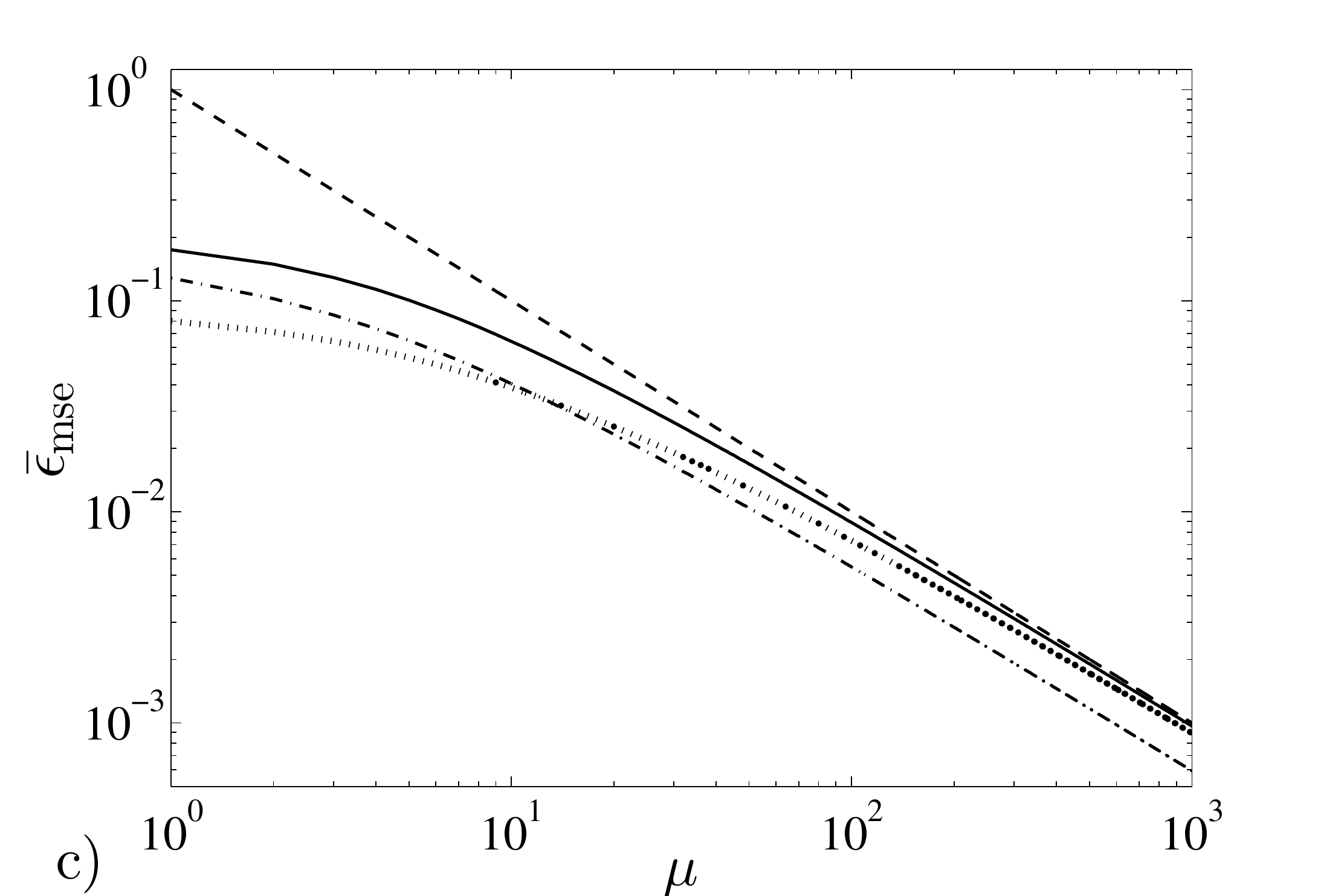}\includegraphics[trim={0.4cm 0cm 0.8cm 0.4cm},clip,width=9.2cm]{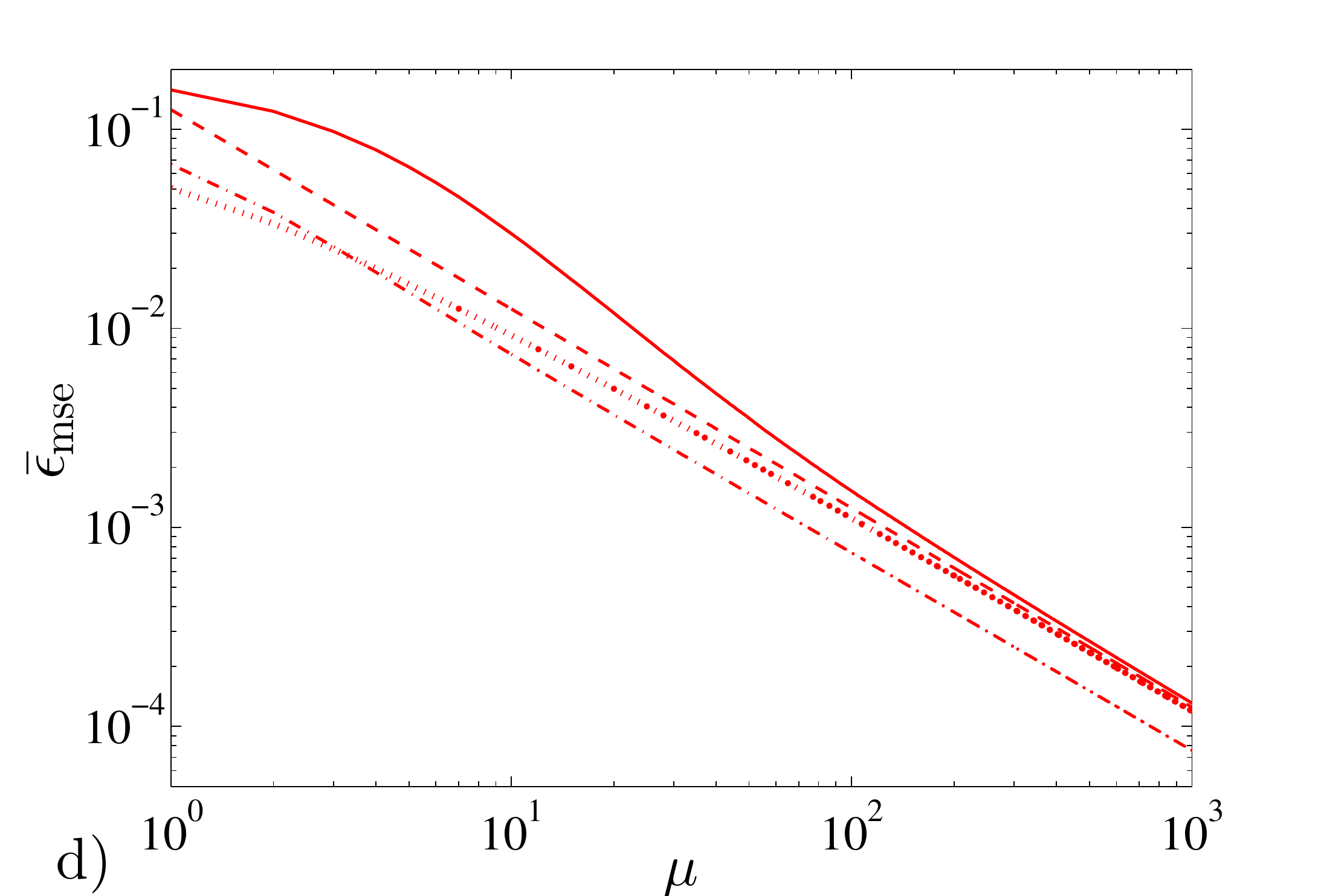}
\includegraphics[trim={0.4cm 0cm 0.8cm 0.4cm},clip,width=9.2cm]{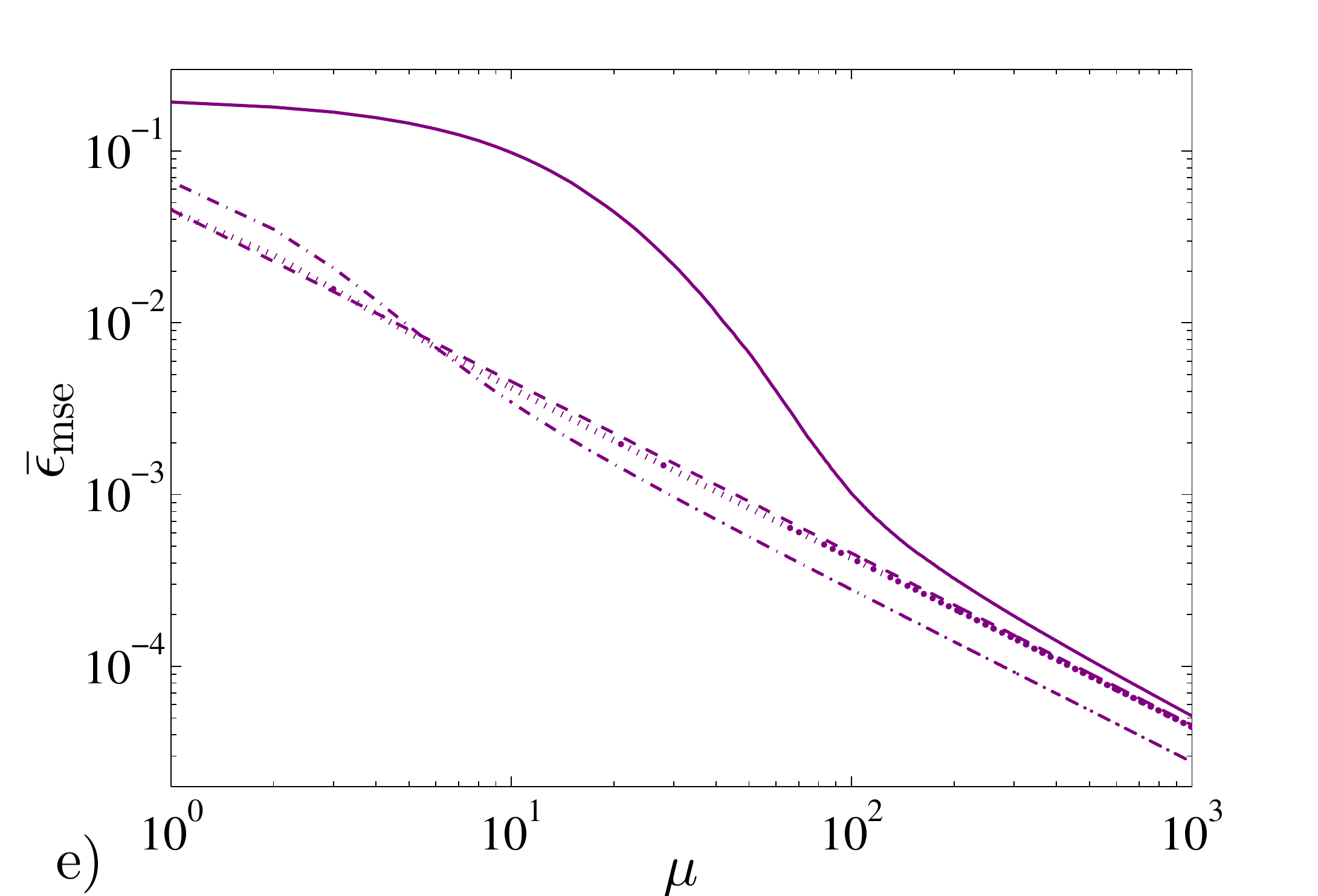}
\caption{Optimal mean square error (solid line), quantum Cram\'{e}r-Rao bound (dashed line), quantum Ziv-Zakai bound (dash-dot line) and quantum Weiss-Weinstein bound (dotted line) for: a) coherent state with $\bar{n} = 2$ and $W_{\mathrm{int}} = \pi$, b) NOON state with $\bar{n} = 2$ and $W_{\mathrm{int}} = \pi/2$, c) NOON state with $\bar{n} = 1$ and $W_{\mathrm{int}} = \pi/2$, d) twin squeezed vacuum with $\bar{n} = 2$ and $W_{\mathrm{int}} = \pi/2$, and e) squeezed entangled state with $\bar{n} = 2$ and $W_{\mathrm{int}} = \pi/2$. This shows that the alternative bounds are valid for any $\mu$. Interestingly, the Ziv-Zakai bound is tighter when $\mu \sim 1$, although the best choice in the asymptotic regime is the Weiss-Weinstein bound. In addition, the Weiss-Weinstein bound and the Cram\'{e}r-Rao bound overlap for the squeezed entangled state, although they are different in the low observation number limit of the other probes.}
\label{consistency}
\end{figure*}

According to Eq.~\ref{fundamental}, the number of required observations will increase when the Fisher information is fixed and the prior knowledge is improved, which is consistent with the results of Figure~\ref{mainresult}. Furthermore, we have seen that the prior width cannot be arbitrarily large if we want to employ certain states in an experiment. Thus, if we maximize the Fisher information at the expense of decreasing the maximum prior uncertainty, and the latter phenomenon is faster, then the number of observations will tend to infinity \footnote{It is important to note that Eq.~\ref{fundamental} only helps to predict cases where $\mu(\rho)$ grows indefinitely. Any other finite result will constitute a necessary but not sufficient condition that the value of the number of observations needed to reach the asymptotic regime must satisfy.}.

This is precisely the case of the family of one-mode states
\begin{equation}
\ket{\psi_0} = \sqrt{1 - \delta}\ket{0} + \sqrt{\delta}\ket{N/\delta}
\label{infinite}
\end{equation}
that was considered in \cite{Alfredo2017}, where $0 < \delta < 1$, $N = \bar{n}$ and $N/\delta$ is an integer. To see it, we notice that the analysis of its periodicity for the unitary transformation $U(\theta) = \mathrm{exp}[-i (a^\dagger a) \theta ]$ indicates that $W_{\mathrm{int}} \leqslant 2\pi\delta/\bar{n}$, which implies that $\Delta \theta^2_p \leqslant \pi^2 \delta^2/(3\bar{n}^2)$, and the quantum Fisher information is $F_q = 4\bar{n}^2(1-\delta)/\delta$. Hence, we have that
\begin{equation}
\mu(\delta) > \frac{3}{4\pi^2 \delta (1- \delta)}.
\label{infinitesolution}
\end{equation}
The Fisher information suggests that we can get an infinite precision in the limit $\delta \rightarrow 0$ for a fixed number of resources per observation $\bar{n}$, but Eq.~\ref{infinitesolution} shows that this conclusion only holds if the total number of resources is actually infinite, which is consistent with the results of \cite{tsang2012, berry2012infinite}. From a physical point of view we conclude that it is not advantageous to use states for which the majority of our resources have to be employed in making our scheme as sensitive as the prior uncertainty that we already had.

To implement the last step that verifies the consistency of our numerical strategy, we need to calculate the alternative bounds that were introduced in Eq.~\ref{qzzb} and Eq.~\ref{qwwb}. Figure~\ref{consistency} shows the results of this procedure. As we expected, both the quantum Ziv-Zakai and Weiss-Weinstein bounds are lower than the numerical mean square error, including the regions where the quantum Cram\'{e}r-Rao bound fails. The reason is that these bounds are valid for both biased and unbiased estimators \cite{tsang2012, tsang2016,  bayesbounds2007}, and as such they correctly lower-bound the uncertainty for low values of $\mu$, in contrast to the Cram\'{e}r-Rao bound. Moreover, the Weiss-Weinstein bound is tight when $\mu \gg 1$, as proven in \cite{tsang2016}. However, its rate of convergence is different from the exact rate obtained in Figure~\ref{mainresult}.b and \ref{mainresult}.d, and the Ziv-Zakai bound is not perfectly tight in any regime. This justifies the use of the direct calculation of the mean square error as a more suitable strategy for this problem. 

\section{Conclusions}

We have explored the limitations of approximating the Bayesian mean square error by the quantum Cram\'{e}r-Rao bound for practical scenarios that are relevant in quantum metrology. This study has been performed by simulating and calculating the mean square error exactly, a process that involves an analysis of the prior knowledge required by a given state and that provides an estimation for the number of observations that are needed to reach the asymptotic regime. Furthermore, we have shown that these results are consistent with the quantum Ziv-Zakai and Weiss-Weinstein bounds, which are always valid. This has allowed us to improve our understanding of both the non-asymptotic regime and the impact of the deviations that the asymptotic theory introduces in the overall performance.

We have applied this strategy to coherent, NOON, twin squeezed vacuum and squeezed entangled states for the estimation of phase shifts in optical interferometry, verifying that the conditions for approaching the Cram\'{e}r-Rao bound crucially vary with the state of the system. Moreover, we have proposed a simple criterion to detect states whose required number of observations is infinite.

From the results of our simulations we can conclude that maximizing the Fisher information alone is not always enough to find the best precision in general. For instance, while a twin squeezed vacuum outperforms NOON states according to the Fisher information, we have found that this conclusion does not hold when the number of observations is low. Similarly, a squeezed entangled state is asymptotically better than the previous examples, but it is the worst choice for small values of $\mu$. In fact, a coherent state with no correlations and a NOON state with less photons per observation outperform it when $\mu \sim 10$. An additional lesson extracted from Section \ref{results} is that future work should revisit the role of inter-mode and intra-more correlations and the use of states with an indefinite number of quanta to enhance the precision in the non-asymptotic regime.

As a consequence, for a real experiment either we need to perform a fully Bayesian analysis or we must estimate explicitly the number of observations that are required to guarantee that we are operating in the asymptotic regime if we want to follow the path of the Fisher information. This practice will improve the quality and fairness of the comparisons between states, helping us to understand the fundamental limits of estimation theory and aiding the design quantum sensing protocols for quantum technologies. 

\section{Acknowledgements}

We acknowledge helpful discussions with Simon Haine. This work was funded by the South East Physics Network
(SEPnet); the United Kingdom EPSRC through the Quantum Technology Hub: Networked Quantum Information Technology (grant reference EP/M013243/1); and the Foundational Questions Institute under the Physics of the Observer Programme (grant no. FQXi-RFP-1601).


\bibliographystyle{apsrev4-1}

\appendix

\section{Quadratic error as an approximation for a periodic error function\label{circular}}

A good experiment should be arranged such that the uncertainty $\bar{\epsilon}$ decreases as a function of the number of observations $\mu$. In that case, the greatest value that $\bar{\epsilon}$ acquires is given by 
\begin{equation}
\bar{\epsilon}|_{\mu=0} = 4\int d\theta p(\theta)\mathrm{sin}^2\left(\frac{g - \theta}{2} \right),
\label{prior_sinerr}
\end{equation}
which is the prior uncertainty for the periodic error function of Eq.~\ref{sinerr} evaluated at $\mu = 0$. Using the uniform prior 
\begin{align}
p(\theta) = 1/(b-a),~\mathrm{for}~ \theta \in [a,b],
\label{prior_asy}
\end{align}
and $p(\theta) = 0$ otherwise, with $a=0$ and $b = W_0$, Eq.~\ref{prior_sinerr} is simplified as
\begin{equation}
\bar{\epsilon}|_{\mu=0} = \frac{4}{W_0}\int_0^{W_0} d\theta ~\mathrm{sin}^2\left(\frac{g - \theta}{2} \right).
\end{equation}
In addition, the minimum of this equation is achieved when the estimator $g$ satisfies 
\begin{equation}
\mathrm{cos}(g - W_0) = \mathrm{cos}(g),
\end{equation}
and for one period this implies that $g = W_0 /2$. Hence,  
\begin{eqnarray}
\bar{\epsilon}|_{\mu=0} &=& \frac{4}{W_0}\int_0^{W_0} d\theta~ \mathrm{sin}^2\left(\frac{W_0}{4} - \frac{\theta}{2} \right)
\nonumber
\\
&=& 2\left[1 - \frac{2}{W_0}\mathrm{sin}\left( \frac{W_0}{2}\right) \right].
\label{sin_muzero}
\end{eqnarray}
If we now expand Eq.~\ref{sin_muzero} up to second order in $W_0$, we find that
\begin{equation}
\bar{\epsilon}|_{\mu=0} \approx \frac{{W_0}^2}{12},
\label{mse_muzero}
\end{equation}  
which is the prior uncertainty that we would have found using the mean square error directly. 

\begin{figure}[t]
\centering
\includegraphics[trim={1.35cm 0.1cm 1.5cm 0.5cm},clip,width=8.9cm]{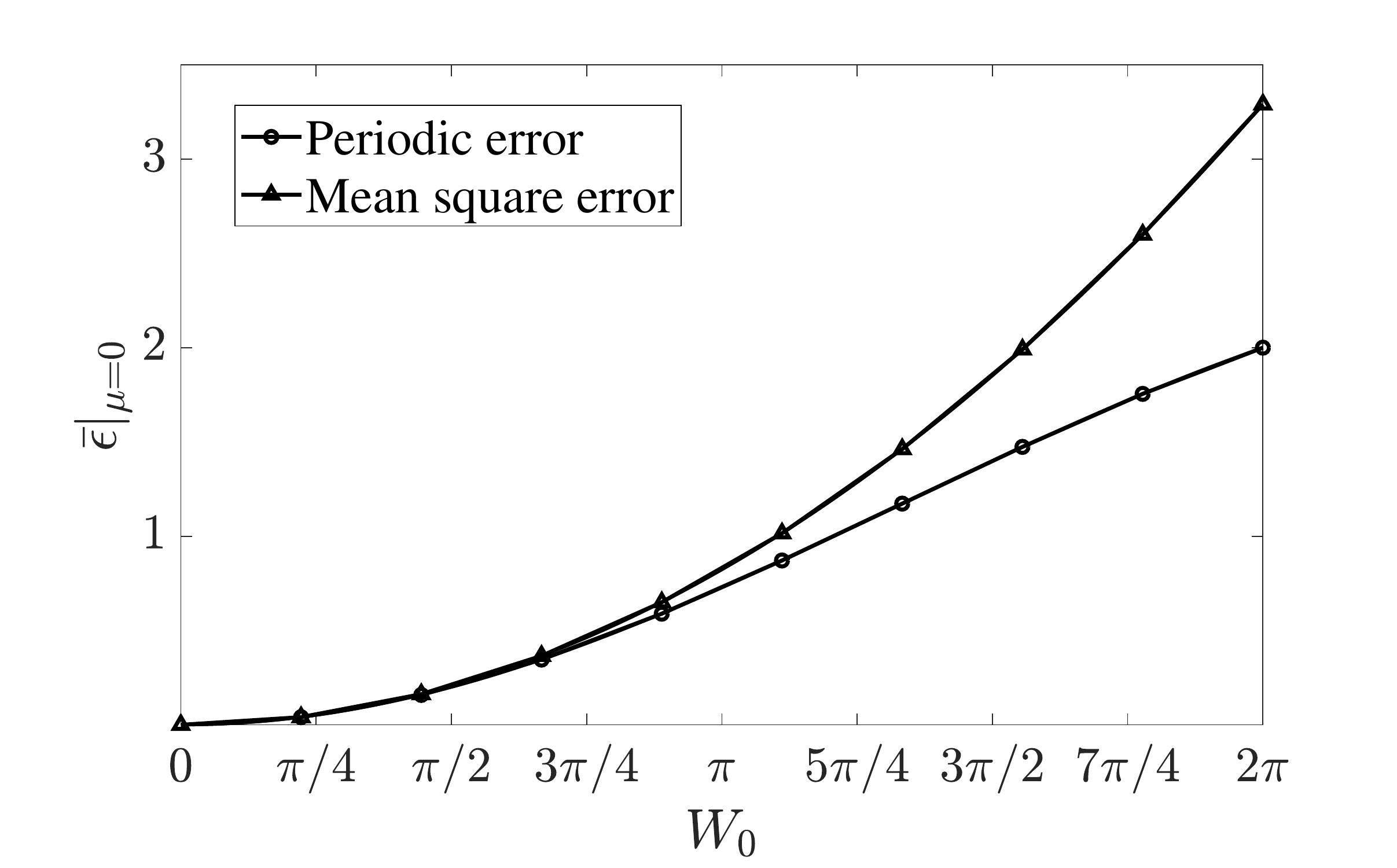}
	\caption{Comparison between the prior uncertainty ($\mu = 0$) given by a periodic error function and that associated to the mean square error as a function of $W_0$. Most of our results in Section \ref{results} are calculated using the values $W_0 = \pi/2$ and $W_0 = \pi/3$.}
\label{circular_mse}
\end{figure}

In Section \ref{results} we calculated the mean square error for NOON, twin squeezed vacuum and squeezed entangled states with $W_0=\pi/2$, and $W_0=\pi/3$ was also employed with both the previous states and for a coherent beam. According to Figure~\ref{circular_mse}, which compares Eq.~\ref{sin_muzero} and Eq.~\ref{mse_muzero} as a function the width $W_0$, the approximation is reasonable for these configurations when $\mu = 0$. Moreover, $\abs{g(\boldsymbol{n}) - \theta}$ will not be greater than $W_0$ for $\mu > 0$, and therefore a similar reasoning can be applied to Eq.~\ref{errwork}. The only scheme for which this approximation is cruder is a coherent state with $W_0 = \pi$.

As a consequence, overall we can conclude that the results of Section \ref{results} are a reasonable numerical approximation to those that we would have obtained should we have used the periodic error function instead, and they certainly constitute an improvement with respect to the usual asymptotic theory. Future work should provide an exact analysis of the non-asymptotic regime for phase estimation.

\section{Asymptotic mean square error}\label{heuristic}

Since the main purpose of this work is to investigate the failure of the Cram\'{e}r-Rao bound for specific scenarios that arise in practice, it is important to keep in mind an intuitive idea about the nature of the approximation that Eq.~\ref{crb} involves. For that reason, we review here the known heuristic argument discussed in Section \ref{approx_classical} using the methods employed in \cite{Jaynes2003, cox2000}. 

Assuming that $p(\boldsymbol{n}|\theta)$ as a function of $\theta$ becomes narrower and concentrated around a unique absolute maximum $\theta_{\boldsymbol{n}}$ when $\mu \gg 1$ \cite{cox2000}, where the observations $\boldsymbol{n}$ were originated from an unknown parameter $\theta'$, and expressing the likelihood as $p(\boldsymbol{n}|\theta) = \mathrm{exp}{\left\lbrace \mathrm{log} \left[p(\boldsymbol{n}|\theta)\right]\right\rbrace}$ in that region, then the first step is to calculate the Taylor expansion
\begin{eqnarray}
\mathrm{log} \left[p(\boldsymbol{n}|\theta)\right] &\approx & \mathrm{log} \left[p(\boldsymbol{n}|\theta_{\boldsymbol{n}})\right]
\nonumber
\\
&+& \frac{1}{2} \frac{\partial^2 \mathrm{log} \left[p(\boldsymbol{n}|\theta_{\boldsymbol{n}})\right]}{\partial \theta^2} (\theta - \theta_{\boldsymbol{n}})^2,
\end{eqnarray}
where the first order term has vanished because $\theta_{\boldsymbol{n}}$ represents a maximum.

Additionally, by the law of large numbers
\begin{eqnarray}
\frac{\partial^2\mathrm{log} \left[p(\boldsymbol{n}|\theta_{\boldsymbol{n}})\right]}{\partial \theta^2} &=& \sum_{i=1}^{\mu} \frac{\partial^2 \mathrm{log} \left[p(n_i|\theta_{\boldsymbol{n}})\right]}{\partial \theta^2}
\nonumber \\
&\approx & \mu\int dn p(n|\theta')\frac{\partial^2 \mathrm{log} \left[p(n|\theta')\right]}{\partial \theta^2},
\label{lln}
\end{eqnarray}
and therefore
\begin{align}
p(\boldsymbol{n}|\theta) \approx \mathrm{exp} \left\lbrace\mathrm{log} \left[p(\boldsymbol{n}|\theta_{\boldsymbol{n}})\right] -\frac{\mu F(\theta_{\boldsymbol{n}})}{2}(\theta - \theta_{\boldsymbol{n}})^2 \right\rbrace,
\label{gaussian_likelihood}
\end{align}
where $F(\theta_{\boldsymbol{n}})$ is the classical Fisher information that arises from expanding the derivative of Eq.~\ref{lln}. 

On the other hand, $\theta_{\boldsymbol{n}} \approx \theta'$ in this case due to the consistency of the maximum of the likelihood \cite{cox2000, kay1993}. Thus, Eq.~\ref{gaussian_likelihood} becomes
\begin{align}
p(\boldsymbol{n}|\theta) \approx p(\boldsymbol{n}|\theta')~ \mathrm{exp} \left\lbrace -\frac{\mu F(\theta')}{2}(\theta - \theta')^2 \right\rbrace.
\label{gaussian_likelihood2}
\end{align}

To obtain the posterior probability defined by Eq.~\ref{bayes} we will use the uniform prior of Eq.~\ref{prior_asy}, understanding that $(b-a)$ is the region where $\theta_{\boldsymbol{n}}$ is unique (see Section \ref{experiment_prior}). Then, we can perform the calculation
\begin{eqnarray}
\int_{a}^b d\theta p(\theta)p(\boldsymbol{n}|\theta)  &\approx & \frac{p(\boldsymbol{n}|\theta')}{b-a}\int_{-\infty}^\infty d\theta \mathrm{e}^{ - \frac{\mu F(\theta')}{2}(\theta - \theta')^2}
\nonumber
\\ 
&=& \frac{p(\boldsymbol{n}|\theta')}{b-a} \sqrt{\frac{2 \pi}{\mu F(\theta')}},
\label{norm_asy}
\end{eqnarray}
where the approximation of the infinite limits holds due to the concentration of $p(\boldsymbol{n}|\theta)$ around a single point. By substituting Eq.~\ref{prior_asy} and Eq.~\ref{norm_asy} into Eq.~\ref{bayes} we recover the Gaussian asymptotic posterior introduced in Eq.~\ref{gaussian}.

Next we need to calculate the variance of the posterior, a step that involves introducing the Gaussian integrals
\begin{eqnarray}
\int_{a}^b d\theta p(\theta|\boldsymbol{n})\theta &\approx & \sqrt{\frac{\mu F(\theta')}{2 \pi}}\int_{-\infty}^\infty d\theta \mathrm{e}^{ - \frac{\mu F(\theta')}{2}(\theta - \theta')^2}\theta
\nonumber
\\ 
&=& \theta',
\nonumber
\\
\int_{a}^b d\theta p(\theta|\boldsymbol{n})\theta^2 &\approx & \sqrt{\frac{\mu F(\theta')}{2 \pi}}\int_{-\infty}^\infty d\theta \mathrm{e}^{ - \frac{\mu F(\theta')}{2}(\theta - \theta')^2}\theta^2
\nonumber
\\ 
&=& (\theta')^2 + \frac{1}{\mu F(\theta')}
\end{eqnarray}
in Eq.~\ref{errobayes}, and we arrive to
\begin{equation}
\epsilon(\boldsymbol{n}) \approx \frac{1}{\mu F(\theta')}.
\label{approx_variance}
\end{equation}
Finally, we notice that the states employed in this work satisfy $F(\theta) = F$ for all $\theta$. Combining this fact with both Eq.~\ref{approx_variance} and $\int d\boldsymbol{n} p(\boldsymbol{n}) = 1$ we conclude that the optimal mean square error in Eq.~\ref{erropt} can be approximated by the Cram\'{e}r-Rao bound in Eq.~\ref{crb}.

\bibliography{arxiv_references_01122017}

\end{document}